\documentclass[prd,aps,twocolumn,floatfix]{revtex4}

\usepackage{amssymb,graphicx,color}

\begin{document}

\title{Spherical excision for moving black holes and\\ 
summation by parts for axisymmetric systems}

\author{Gioel Calabrese and David Neilsen}

\affiliation{Department of Physics and Astronomy, Louisiana State
University, 202 Nicholson Hall, Baton Rouge, Louisiana 70803-4001}

\date{\today}

\begin{abstract}
It is expected that the realization of a convergent and long-term
stable numerical code for the simulation of a black hole inspiral
collision will depend greatly upon the construction of stable
algorithms capable of handling smooth and, most likely, time dependent
boundaries.  After deriving single grid, energy conserving
discretizations for axisymmetric systems containing the axis
of symmetry, we present a new excision method for moving black holes
using multiple overlapping coordinate patches, such that each boundary
is fixed with respect to at least one coordinate system.  This
multiple coordinate structure eliminates all need for extrapolation, a
commonly used procedure for moving boundaries in numerical relativity.

We demonstrate this excision method by evolving a massless
Klein-Gordon scalar field around a boosted Schwarzschild black hole in
axisymmetry.  The excision boundary is defined by a spherical
coordinate system co-moving with the black hole.
Our numerical experiments indicate that arbitrarily high boost
velocities can be used without observing any sign of instability.

\end{abstract}

\maketitle

\section{Introduction}

Inspiraling black holes are among the strongest astrophysical sources
of gravitational radiation.  The expectation that such systems may
soon be studied with gravitational wave detectors has focused
attention on solving Einstein's equations for predictions of
gravitational wave content.  Although the Einstein equations present
several unique challenges to the numerical relativist~\cite{L}, on
several of which we do not elaborate here, black holes present one
particular additional challenge: they contain physical curvature
singularities.  While the infinities of the gravitational fields
associated with this singularity cannot be represented directly on a
computer, the spacetime near the black hole must be given adequately
to preserve the proper physics.  

Different strategies have thus been developed to computationally
represent black holes, while removing the singularity from the grid.
One method exploits the gauge freedom of general relativity
by choosing a time coordinate that advances normally far from a
singularity, slows down as a singularity is approached, and freezes in
the immediate vicinity.  Coordinates with this property are ``singularity
avoiding''~\cite{SmaYor,EarSma,BarPir}.  While singularity avoiding
coordinates have some advantages, one potential disadvantage is that
the hypersurfaces of constant time may become highly distorted, leading
to large gradients in the metric components. 
These slice-stretching (or ``grid-stretching'') effects,
however, can be partially avoided through an advantageous combination of
lapse and shift conditions.  For example, long-term evolutions of single
black holes have been reported by Alcubierre et al.~\cite{AEI}.
Singularity avoiding slicings may be combined with black hole excision, a 
second method for removing the singularities from the computational 
domain.  Currently, long-term binary black hole evolutions have only been
performed using both techniques together.

Excision is based on the physical properties of event horizons and the
expectation that singularities always form within such horizons, and
thus cannot be seen by distant observers, as formulated by the Cosmic
Censor Conjecture~\cite{W}.  As no future-directed causal curve
connects events inside the black hole to events outside, Unruh
proposed that one could simply remove the black hole from the
computational domain, leaving the exterior computation
unaffected~\cite{Unr}.  Thus the black hole singularity is removed by
placing an inner boundary on the computational domain at or within the
event horizon.  Excision has been extensively used in numerical
relativity in the context of Cauchy formulations
\cite{SeiSue,SchShaTeu,BBHGCA,Brandt,AlcBru,YoBauSha,KidSchTeu,Pre,%
ShoSmiSpeLagSchFis, CalLehNeiPulReuSarTig,SpeSmiKelLagSho}.  In
particular, excision with moving boundaries, which is the primary
focus of this paper, was explored
in~\cite{BBHGCA,Brandt,YoBauSha,ShoSmiSpeLagSchFis,SpeSmiKelLagSho}.

The physical principles that form the basis of excision make the idea
beautiful in its simplicity. Translating them into a workable
numerical recipe for black hole evolutions, on the other hand,
requires some attention to detail.  Two general questions arise
regarding the implementation of excision, (1) Where and how to define
the inner boundary?  and (2) How to move the boundary?  The first
question applies to all excision algorithms, while the last question
is specific to implementations where the excision boundary moves with
respect to the grid.  In addressing these questions we assume a
symmetric (or at least strongly) hyperbolic formulation~\cite{R}.
This is because excision fundamentally relies on the characteristic
structure of the Einstein equations near event horizons, a structure
which can only be completely defined and understood for strongly and
symmetric hyperbolic sets of equations.

The first question involves several considerations, including the
location of the boundary, its geometry, and its discrete
representation.  The requirement that all modes at the excision
boundary are leaving the computational domain can be non trivial.  It
may appear that this condition would be satisfied simply by choosing
any boundary within the event horizon (or, for practical purposes, the
apparent horizon).  However, the outflow property of the excision
boundary depends on the characteristic speeds of the system in the
normal directions to the boundary.  For example, in the analytic
Schwarzschild solution, assuming that the system has characteristic
speeds bounded by the light cone, a spherical boundary can be excised
at $r \le 2M$.  A cubical boundary, on the other hand, imposes an
onerous restriction on the excision volume: in Cartesian Kerr--Schild
coordinates the faces of a cube centered on the black hole must be
less than $4\sqrt{3}/9M \approx 0.7698M$ in
length~\cite{Sch,CalLehNeiPulReuSarTig}.  Remarkably, as was first
noticed by Lehner~\cite{Leh}, for the rotating Kerr solution in
Kerr--Schild coordinates a well-defined cubical excision boundary is
impossible for interesting values of the spin parameter.  (See the
appendix for further discussion.)  Whereas with a pseudospectral
collocation method the implementation of a smooth spherical excision
boundary is trivial~\cite{PfeKidSchTeu}, this is generally not the case
for finite differencing.  As may be expected, smooth boundaries, which
can be adapted to the spacetime geometry, allow the excision boundary
to be as far from the singularity as possible, making the most
efficient use of the technique.

The discrete representation of boundaries can be a delicate issue,
especially in numerical relativity where many large-scale finite difference
computations are done in
Cartesian coordinates.  We focus our attention on smooth boundaries
that may be defined as a constant value in the computational
coordinates, e.g., $r=r_0$ in spherical coordinates describes a simple
spherical boundary.  The importance of accurately
representing smooth boundaries has been demonstrated for the Euler
equations, for example, by Dadone and Grossman for finite volume
methods~\cite{DadGro}, and Bassi and Rebay~\cite{BasReb} using finite
elements.  Bassi and Rebay studied high resolution planar fluid flow
around a cylinder.  They report spurious entropy production near the
cylinder wall, which corrupts the solution even on extremely refined
grids, when the cylindrical boundary is approximated by a polygon.
Furthermore, in the conformal field equations approach to general
relativity, a smooth boundary is required to avoid uncontrollable
numerical constraint violation~\cite{Sascha}.

The second question applies to excision boundaries that move with
respect to the grid.  When the inner boundary moves, points that
previously were excised enter the physical part of the grid, and must
be provided with physical data for all fields.  In recently proposed
excision algorithms, these data are obtained by extrapolating the
solution from the physical domain of the calculation.  Numerical
experiments have indicated that the stability of the method is very
sensitive to the details of the extrapolation, see e.g.,
Ref.~\cite{LehHuqGar,YoBauSha,ShoSmiSpeLagSchFis}.  
To examine the black hole excision problem
with moving inner boundaries, we adopt an approach with some unique
features.  The heart of our method for moving excision is to use
multiple coordinate patches such that each boundary is at a fixed
location in one coordinate system.  Adapting coordinate patches to the
boundary geometry allows us to excise as far from the singularity as
possible and simplifies the determination of the outflow character of
the excision boundary.  The motion of the boundaries is incorporated
through the relationships among the various coordinate systems.  The
grids representing the different coordinate patches overlap and
communicate via interpolation.  This technique is an extension of the
one used in well-posedness proofs for problems in general domains (see
Sec.~13.4 of \cite{GKO-Book}).  In this paper we demonstrate the
algorithm by solving the massless Klein-Gordon equation on a fixed,
boosted Schwarzschild background.  We find that the algorithm is
stable for (apparently) all values of the boost parameter, 
$\beta=v/c$, and present results here showing stable
evolutions for several cases with $\beta \leq 0.95$.

We specialize to axially symmetric spacetimes to reduce the
computational requirements for our single-processor code.  Axially
symmetric spacetimes have sometimes been avoided in numerical
relativity, with notable exceptions, see e.g.,~\cite{Axi}, owing to
the difficulties in developing stable finite difference
equations containing the axis of symmetry.  In this paper we
further present finite differencing methods for the wave 
equation in axially 
symmetric spacetimes in canonical cylindrical and spherical coordinates.  
These differencing schemes are second order accurate
and their stability for a single grid is proved using the energy
method \cite{GKO-Book}.  Maximally dissipative boundary conditions are
applied using the projection method of Olsson \cite{Ols}.  We present
the differencing algorithm in detail, and indicate precisely how
boundary conditions are applied.

This paper is organized as follows: In
Sec.~\ref{Sec:Overview_excision} we motivate our 
approach and review the overlapping grid method.  We recall the
concept of conserved energy for a first order symmetrizable
hyperbolic system in Sec.~\ref{Sec:FOSH} and provide an energy
preserving discretization.  In Sec.~\ref{Sec:WE_flat} we analyze the
axisymmetric wave equation around a Minkowski background as an introduction
to our numerical methods.  The
analysis is then repeated for the black hole background case in
Sec.~\ref{Sec:WE_BH}.  The excision of a boosted black hole with the
overlapping grid method is described in Sec.~\ref{Sec:Boosted}.  The
numerical experiments, along with several convergence tests, are included
in Sec.~\ref{Sec:NumExp}.

\section{Overview}
\label{Sec:Overview_excision}

Our primary goal is to obtain a numerical algorithm for excision with
moving black holes that is stable and convergent (in the limit that
the mesh spacing goes to zero).  These desired properties for the
discrete system closely mirror the continuum properties of well-posed
initial boundary value problems (IBVPs): the existence of a unique
solution that depends continuously on the initial and boundary data.
Furthermore, we believe that we will not obtain long-term convergent
discrete solutions {\it unless} the underlying continuum problem is
also well-posed.  Unfortunately there are few mathematical results
concerning the well-posedness of general classes of equations.  The
energy method, however, can be used with symmetric hyperbolic IBVPs,
and gives sufficient conditions for well-posedness.

\begin{figure}[t]
\begin{center}
\includegraphics*[height=8cm]{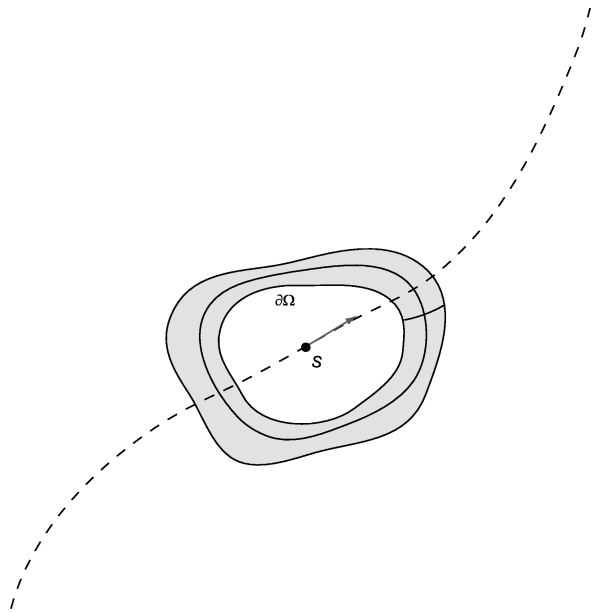}
\caption{A singularity $S$ surrounded by an event horizon $\partial
\Omega$ is moving with respect to the base coordinate system.  A
coordinate patch (shaded region) adapted to $\partial\Omega$ follows
the motion of the singularity.  With respect to this patch, $\partial
\Omega$ is a purely outflow boundary and requires no boundary
conditions.  The base system terminates somewhere inside the shaded
region and it gets boundary data from the moving patch.  Similarly,
the data at the outer boundary of the moving patch is taken from the
base system.}
\label{Fig:patches}
\end{center}
\end{figure}

When a black hole moves with respect to some coordinate system, the
inner excision boundary must also move.  We use multiple coordinate
patches, such that every boundary is fixed with respect to at least
one coordinate system.  Coordinate transformations relate the
coordinate systems, and become time dependent when the hole 
moves.  The movement of the inner boundary is also expressed by these
time-dependent coordinate transformations.  These ideas are
illustrated in Fig.~\ref{Fig:patches}.
In our axially symmetric model problem of a scalar field on a boosted
Schwarzschild spacetime, the computational frame is covered with
cylindrical coordinates, while a second patch of spherical coordinates
is co-moving with the black hole.  (In these coordinates the event
horizon is always located at $r=2M$ while the time coordinate is taken
from the cylindrical patch so that data on all grids are
simultaneous.)  The inner boundary of the spherical grid, located at
or within the event horizon, is a simple outflow boundary, and
requires no boundary condition.  The cylindrical domain has an inner
boundary somewhere near the black hole, whether inside or outside of
the horizon is immaterial, as long as it is covered by the spherical
coordinate patch.  An exchange of information between the two
coordinate patches is required to provide boundary conditions at the
inner cylindrical boundary and the outer spherical boundary.

On each grid the discrete system is constructed using the energy
method~\cite{GKO-Book}.  We define an energy for the semi-discrete
system and, using difference operators that satisfy summation by
parts, we obtain a discrete energy
estimate~\cite{CalLehNeiPulReuSarTig}.  Well-posed boundary conditions
can then be identified by controlling the boundary terms of the
discrete energy estimate.  The conditions are discretized using
Olsson's projection method~\cite{Ols}.  In particular, the symmetry
axis ($\rho=0$ in canonical cylindrical coordinates) is included in
the discrete energy estimate~\cite{Sar}, allowing us to naturally
obtain a stable discretization for axisymmetric systems.

We implement our excision algorithm using overlapping grids, also
known as composite mesh difference method~\cite{GKO-Book,Sta,CheHen}.
The two grids are coupled by interpolation, which is done for all the
components of the fields being evolved.  If the system is hyperbolic
this means that one is actually over specifying the problem.  However,
as it is pointed out in Sec.~13.4 of \cite{GKO-Book} and as it is
confirmed by our experiments, this does not lead to a
numerical instability.  The fully discretized system is completed by
integrating the semi-discrete equations with an appropriate method for
ODEs; we choose third and fourth order Runge-Kutta, which does not
spoil the energy estimate of the semi-discrete
system~\cite{CalLehNeiPulReuSarTig}.  Kreiss-Oliger
dissipation~\cite{KreOli} is added to the scheme, as some explicit
dissipation is generally necessary for stability with overlapping
grids~\cite{OlsPet}.  Whereas the stability theory for overlapping
grids for elliptic problems is well developed, there are very few
results concerning hyperbolic systems.  Starius presents a stability
proof for overlapping grids in one dimension~\cite{Sta}.  
Finally, we note that
Thornburg has also explored multiple grids in the context of numerical
relativity with black hole excision~\cite{ThoI}.

The structure of the overlapping grids used in this work is
illustrated in Fig.~\ref{Fig:Overlapping}.  The additional
complication of the axis of symmetry is discussed below.  For
simplicity we choose the outer boundary to be of rectangular shape.
The introduction of a smooth spherical outer boundary, along with
another grid overlapping with the base cylindrical grid, is certainly
possible and, we believe, likely to improve the absorbing character of
the outer boundary when the incoming fields are set to zero.

\section{The wave equation}
\label{Sec:FOSH}

To demonstrate our excision algorithm, we choose the evolution of a
massless Klein-Gordon scalar field on an axisymmetric, boosted
Schwarzschild background as a model problem.  In this section we
summarize basic definitions for linear, first order hyperbolic
initial-boundary value problems~\cite{GKO-Book,R}.  We employ the
energy method to identify well-posed boundary conditions.  The
discrete version of this method, based on difference operators
satisfying the summation by parts rule~\cite{Str}, is then used to
discretize the right hand side of the system and the boundary
conditions on a single rectangular grid.  (For an introduction to
these methods in the context of numerical relativity see
Refs.~\cite{CalLehNeiPulReuSarTig,LongPaper,LongPaperII}.)  We then
introduce the axisymmetric scalar field equations on a curved
background, along with their semi-discrete approximation.

In this paper we adopt the Einstein summation convention and
geometrized units ($G=c=1$).  Latin indices range over the
spatial dimensions, and Greek indices label spacetime components.

\subsection{Hyperbolic systems in first order form}

Consider a linear, first order, hyperbolic IBVP in two spatial dimensions,
consisting of a system of partial differential evolution equations,
and initial and boundary data, of the form
\begin{eqnarray}
&&\partial_t u = A^i(t,\vec{x}) \partial_{i} u +
B(t,\vec{x}) u\quad (t,\vec{x}) \in [0,T]\times \Omega
\label{Eq:FOSH}\\
&&u(0,\vec{x}) = f(\vec{x})\qquad \vec{x} \in \Omega
\label{Eq:initialdata}\\
&&Lu(t,\vec{x}) = g(t,\vec{x}) \qquad (t,\vec{x}) \in [0,T]\times
\partial\Omega,
\label{Eq:boundarydata}
\end{eqnarray}
where $i=1,2$, $u=u(t,\vec{x})$ and $f(\vec{x})$ are vector valued
functions with $m$ components, $A^i$ and $B$ are $m\times m$ matrices
that depend on the spacetime coordinates but not on the solution $u$,
and $\partial_i$ stands for $\partial / \partial_{x^i}$.  The boundary of $\Omega
\subset \mathbb{R}^2$ is assumed to be a simple smooth curve.  The
operator $L$ and the data $g$ that appear in the boundary
condition (\ref{Eq:boundarydata}) will be defined below in
Eqs.~(\ref{Eq:maxdissip}--\ref{Eq:L}).

\subsubsection{Strong and symmetric hyperbolicity}

System (\ref{Eq:FOSH})--(\ref{Eq:boundarydata}) is said to be {\em
strongly} hyperbolic in $O\subset [0,T]\times\Omega$ if, at each point
$(t_0,\vec{x}_0)\in O$, the matrix
\begin{equation}
\hat{P}(t_0,\vec{x}_0,\vec{\omega}) =  A^j(t_0,\vec{x}_0)\omega_j\,,
\end{equation}
with $\vec{\omega} \in \mathbb{R}^2$ and $| \vec{\omega}|^2 =
\omega_1^2 +\omega_2^2 = 1$, can brought into real diagonal form by a
transformation $T(\vec{\omega})$, such that $T(\vec{\omega})$ and
$T^{-1}(\vec{\omega})$ are uniformly bounded with respect to
$\vec\omega$.  The system is said to be {\em symmetric} or {\em
symmetrizable} hyperbolic in $O$ if, at each point $(t_0,\vec{x}_0)\in
O$, there exists a smooth, symmetric positive definite matrix
$H(t_0,\vec{x}_0)$, independent of $\vec\omega$, such that $HA^i =
(HA^i)^T$ for $i=1,2$.  The matrix $H$ is usually called the {\em
symmetrizer}.  Clearly, a symmetric hyperbolic system is also strongly
hyperbolic.  Strong hyperbolicity is a necessary condition for
well-posedness and consequently for the construction of stable
numerical schemes.

\subsubsection{Characteristic speeds}

The {\em characteristic speeds} in the
direction $\vec n = (n_1,n_2) \in \mathbb{R}^2$, with $n_1^2+n_2^2 =
1$, at the point $(t_0,\vec{x}_0)\in [0,T]\times\Omega$ are the
eigenvalues of $A^n(t_0,\vec{x}_0)\equiv n_i
A^i(t_0,\vec{x}_0)$.
In Sec.~\ref{Sec:NumExp} we will show how the maximum value of the
characteristic speeds in the region $[0,T]\times\Omega$ can be used to
compute an upper bound for the ratio between the time step and the
spatial mesh size.  

\subsubsection{Energy method}

The specification of proper boundary conditions requires careful
consideration in order to achieve a well-posed
IBVP, and we use the energy method to identify appropriate boundary
conditions~\cite{KL-Book,GKO-Book}.  Here one defines the
energy of the system at time $t$ to be
\begin{equation}
E(t) = \| u(t,\cdot ) \|_H^2 =\int_{\Omega} u^T(t,\vec x) H(t,\vec x)
u(t,\vec x) \,d^2
x\,,
\label{Eq:energy}
\end{equation}
where $H$ is some positive definite $m\times m$ symmetric matrix and
$u^T$ denotes the transpose of $u$.  To ensure
continuous dependence of the solution on the initial and boundary
data, the energy must be bounded in terms of appropriate norms of the
data.  To determine this bound one usually takes a time derivative of
the energy (\ref{Eq:energy}), with the further assumptions that $u$ is
a smooth solution of (\ref{Eq:FOSH}) and that $H$ is a symmetrizer.
The energy estimate is then
\begin{eqnarray}
&&\frac{d}{dt} E(t) = \int_{\partial\Omega} u^T HA^n u \,
d \sigma\label{Eq:E_dot}\\
&&+\int_{\Omega} u^T \left( \partial_t H + HB + (HB)^T -
\partial_i (HA^i)\right) u \, d^2 x\,,\nonumber
\end{eqnarray}
where Gauss' theorem was used to obtain the boundary term, and $n_i$ is the
outward unit normal to the boundary $\partial\Omega$.  
To control the growth of the energy of the solution, we naturally need to
control both the boundary and volume integrals.  
We consider the boundary integral first.  

The matrix $HA^n$ is symmetric, and can be brought into diagonal form
by an orthogonal transformation $Q(n)$,
\begin{equation}
Q^T(n) HA^n Q(n) = \Lambda = {\rm diag} (\Lambda_+, -\Lambda_-, 0)\,,
\end{equation}
where $\Lambda_{\pm} > 0$ are positive definite diagonal matrices, the
eigenvalues of which, in general, do not coincide with the
characteristic speeds.  This allows one to rewrite the integrand of
the boundary integral in Eq.~(\ref{Eq:E_dot}) by introducing the
vector $w^{(n)} =
({w^{(+\Lambda_+;n)}},{w^{(-\Lambda_-;n)}},w^{(0;n)})^T = Q^T(n) u$ as
the difference between two non-negative terms,
\begin{eqnarray}
&&u^T HA^n u =\\
 && {w^{(+\Lambda_+;n)}}^T \Lambda_+
{w^{(+\Lambda_+;n)}} - {w^{(-\Lambda_-;n)}}^T \Lambda_-
{w^{(-\Lambda_-;n)}}\,.\nonumber
\end{eqnarray}
The components of $w^{(n)}$ are the {\em characteristic variables}
in the direction $\vec n$.  In particular, the components of
$w^{(+\Lambda_+;n)}$ are the {\em ingoing} characteristic variables,
and the components of $w^{(-\Lambda_-;n)}$ are the {\em outgoing}
characteristic variables.  We see that prescribing
homogeneous boundary conditions (${w^{(+\Lambda_+;n)}} = S
{w^{(-\Lambda_-;n)}}$, with $S$ sufficiently small, i.e.,~$S^T
\Lambda_+S \le \Lambda_-$), ensures that the boundary term will give a
non-positive contribution to the energy estimate.  The $S=0$ case (no
coupling) is of particular interest as it usually yields a good
approximation for absorbing (Sommerfeld) boundary conditions.

The second term of the energy estimate (\ref{Eq:E_dot}), the volume
integral, can be estimated by $2\alpha \| u(t,\cdot)\|^2_H$, where
$\alpha = \frac{1}{2} \max_{(t,\vec x)} \| \partial_t H + HB + (HB)^T
- \partial_i (HA^i)\|$ is a constant that does not depend on the
solution.  Thus, for homogeneous boundary conditions we have
\begin{equation}
\frac{d}{dt} \| u(t,\cdot) \|^2_H \le 2 \alpha \| u(t,\cdot)\|^2_H\, ,
\end{equation}
which implies that $\|u(t,\cdot)\|_H \le \exp(\alpha t) \|f \|_H$.
Similar energy estimates can be obtained for inhomogeneous boundary
conditions~\cite{KL-Book, GKO-Book}, i.e.,
\begin{equation}
w^{(+\Lambda_+;n)} = S w^{(-\Lambda_-;n)} + g\,,
\label{Eq:maxdissip}
\end{equation}
where $g$ has to satisfy compatibility conditions with the initial data.

Boundary conditions of the form (\ref{Eq:maxdissip}) are referred to
as {\em maximally dissipative} boundary conditions \cite{LaxPh}.  From
Eq.~(\ref{Eq:maxdissip}) we see that the operator $L$ introduced in
(\ref{Eq:boundarydata}) has the form
\begin{equation}
L = P^{(+)}Q^T(n) - SP^{(-)}Q^T(n)\, ,
\label{Eq:L}
\end{equation}
where $P^{(+)} (w^{(+)},w^{(-)},w^0)^T = (w^{(+)},0,0)^T$ and $P^{(-)}
(w^{(+)},w^{(-)},w^0)^T = (0,w^{(-)},0)^T$.
Finally, it is important to recognize that if the matrix $H$ is not a
symmetrizer, as would be the case if $H$ symmetrizes $A^i$ but fails
to be positive definite, then the boundary condition above will not,
in general, lead to a well-posed problem, as one would likely end up
specifying boundary data to the wrong quantities.

\subsubsection{Strict stability}

Discretizing the spatial derivatives in the right hand side of system 
(\ref{Eq:FOSH}), but leaving time continuous, leads to a system of 
ODEs called the {\em semi-discrete system}.  If an initial value problem 
satisfies the
estimate $\|u(t,\cdot) \|_H \le K\exp(\alpha t) \| u(0,\cdot) \|_H$ at
the continuum, it would be desirable to obtain a discretization such
that a similar estimate holds at the discrete level.  Following
\cite{GKO-Book}, we will say that a semi-discrete system is {\em
strictly stable} if
\begin{equation}
\| u(t) \|_h \le K_S e^{\alpha_S t} \|u(0)\|_h
\end{equation}
where $\alpha_S \le \alpha + {\cal O}(h)$ and $\|\cdot \|_h$ is a
discrete energy consistent with the one of the continuum.

\subsubsection{Conserved energy}

Clearly, the requirement that $HA^i$ be symmetric, does not
uniquely determine the symmetrizer.  For example, if $H$
is a symmetrizer, then $f H$ with $f>0$ is also a symmetrizer.  
In some circumstances, as for the scalar field considered here,
it is possible to select a preferred symmetrizer which satisfies 
the additional requirement
\begin{equation}
\partial_t H + HB + (HB)^T - \partial_i (HA^i) = 0\,.
\label{Eq:conserved}
\end{equation}
When this condition holds, the energy defined by that symmetrizer will
be conserved.  By this we mean that the change in energy of our system
is solely due to the boundary term of (\ref{Eq:E_dot}), which can be
controlled by using maximal dissipative boundary conditions
(\ref{Eq:maxdissip}). In particular, when homogeneous boundary conditions are
used, or when no boundaries are present, the energy cannot increase.

\subsubsection{Energy conserving schemes}
Let us assume momentarily that that there exists a symmetrizer for
which Eq.~(\ref{Eq:conserved}) holds (the system admits a conserved
energy), and that no boundaries are present.  In the variable
coefficient case (more precisely, if $\partial_j (HA^i) \neq 0$ for
$i=j$), the naive discretization $\partial_t u = A^i D_i u + B u$,
where $u$ now represents a vector valued grid function, although
strictly stable when a second order accurate centered difference
operator is used \cite{Ols}, does not conserve the discrete energy
\begin{equation}
E = (u,Hu)_h = h_1h_2 \sum_{ij} u^T_{ij} H_{ij} u_{ij}\,,\
\label{Eq:discE}
\end{equation}
where $H_{ij} = H(t,\vec{x}_{ij})$.  Its time derivative gives 
\begin{equation}
\frac{d}{dt} E(t) = (u,[HA^i,D_i]u)_h + (u,\partial_i (HA^i)u)_h \neq
0\,,
\label{Eq:naive}
\end{equation}
where we used the fact that $(v,D_iu)_h + (D_iv,u)_h = 0$.  The lack
of a Leibniz rule at the discrete level is only partly responsible for
this.  In general, even if this rule were satisfied, the discrete
estimate would not vanish, $\frac{d}{dt}E = (u,(\partial_i (HA^i) -
D_i (HA^i))u)_h \neq 0$.  Any semi-discrete approximation that
preserves the discrete energy (\ref{Eq:discE}) is an {\em energy
conserving} scheme.  Remarkably, whenever a system admits a conserved
energy at the continuum it is always possible to construct an energy
conserving scheme~\cite{CalLehNeiPulReuSarTig,LongPaper,LongPaperII}.
The following ``$1/2 + 1/2$'' splitting, for example,
\begin{eqnarray}
\partial_t u &=& \frac{1}{2} A^i D_i u + \frac{1}{2} H^{-1} D_i (HA^iu)\label{Eq:splitting}\\
&&+ \left( B- \frac{1}{2} H^{-1} \partial_i (HA^i)\right) u\,,\nonumber
\end{eqnarray}
ensures that the discrete energy (\ref{Eq:discE}) remains constant. Clearly, an energy
conserving scheme is strictly stable, since $\alpha = \alpha_S =0$.
We note that, depending on the problem, there may be alternative,
simpler discretizations than the ``$1/2 + 1/2$'' splitting which lead
to the same energy estimate.
Moreover, a discretization such as (\ref{Eq:splitting}) is a consistent
approximation of $\partial_t u = A^i\partial_i u + Bu$ whether or not
condition (\ref{Eq:conserved}) holds.

\subsubsection{Rectangular grid}

Consider a rectangular domain 
$\Omega = \{ (x^1,x^2)| x^1_{\rm min} \le x^1 \le x^1_{\rm max} , x^2_{\rm min}
\le x^2 \le x^2_{\rm max}\}$,  with the grid points
$\vec{x}_{ij} = (x^1_{\rm min} + i h_1, x^2_{\rm min} + j h_2)$, $i =
0, \ldots, N_1$ and $j = 0, \ldots, N_2$, and $h_k = (x^k_{\rm max} -
x^k_{\rm min})/N_k$, $k = 1, 2$.  From the continuum analysis we
expect that boundary data should be given to the incoming
characteristic variables in the direction orthogonal to the boundary
surface.  In addition, at the corners the boundary data has to satisfy
compatibility conditions.  We now repeat the same analysis for the
semi-discrete system in order to determine appropriate boundary
conditions for the computational grid. In particular, we examine the
application of boundary conditions at the corner points of the grid.

We define the following one dimensional scalar products between vector
valued grid functions,
\begin{equation}
(u,v)_{h_1} = h_1 \sum_{i=0}^{N_1} u^T_{i} v_{i} \sigma_i\, , \qquad 
(u,v)_{h_2} = h_2 \sum_{i=0}^{N_2} u^T_{i} v_{i} \sigma_i\, ,
\end{equation}
where $\sigma_i = \{ 1/2, 1, \ldots, 1, 1/2 \}$.  The 2D scalar
product is
\begin{equation}
(u,v)_h = h_1 h_2 \sum_{i=0}^{N_1}\sum_{j=0}^{N_2} u^T_{ij} v_{ij} 
        \sigma_i \sigma_j \, .
\end{equation}
To simplify the notation we introduce $D^{(i)} = D^{(x^i)}$.  If we
approximate $\partial_1$ with the second order centered difference
operator $D^{(1)}_0u_{ij} = (u_{i+1,j}-u_{i-1,j})/(2h_1)$ in the
interior ($1\le i \le N_1-1$, $0 \le j \le N_2$) and with the first
order one-sided difference operators $D^{(1)}_+ u_{0j} =
(u_{1,j}-u_{0,j})/h_1$, $D^{(1)}_-u_{N_1j} =
(u_{N_1,j}-u_{N_1-1,j})/h_1$ at the $x^1={\rm const.}$ boundary we
have that
\begin{eqnarray}
&&(u,D^{(1)}v)_h + (D^{(1)}u,v)_h\label{Eq:sumbyparts1}\\
&=& h_2 \sum_{j=0}^{N_2} \left(h_1\sum_{i=0}^{N_1} u_{ij}
D^{(1)} v_{ij} \sigma_i + h_1 \sum_{i=0}^{N_1} D^{(1)} u_{ij} v_{ij}
\sigma_i\right) \sigma_j\nonumber\\
&=& (u_{i\cdot},v_{i\cdot})_{h_2}|_{i=0}^{i=N_1}\, . \nonumber
\end{eqnarray}
Similarly, if $D^{(2)}=D_0^{(2)}$ in the interior and $D^{(2)} =
D_{\pm}^{(2)}$ at the $x^2={\rm const.}$ boundary, we have that 
\begin{equation}
(u,D^{(2)}v)_h + (D^{(2)}u,v)_h = \left. (u_{\cdot j},v_{\cdot
j})_{h_1}\right|_{j=0}^{j=N_2}\, .
\label{Eq:sumbyparts2}
\end{equation}

If these simple finite difference operators are used to approximate
the spatial derivatives in, for example, (\ref{Eq:splitting}), the
time derivative of the discrete energy
\begin{equation}
E = (u,Hu)_h = h_1h_2 \sum_{ij} u^T_{ij} H_{ij} u_{ij} \sigma_i \sigma_j,
\end{equation}
gives
\begin{eqnarray}
\frac{d}{dt} E &=& (u_{i\cdot},(HA^1u)_{i\cdot})_{h_2}|_{i=0}^{i=N_1} +
(u_{\cdot j}, (HA^2u)_{\cdot j})_{h_1}|_{j=0}^{j=N_2}\nonumber\\
 &&+ (u, (\partial_t
H + HB +(HB)^T - \partial_i (HA^i)) u)_h \, ,
\label{Eq:goodestimate}
\end{eqnarray}
where we have not assumed energy conservation.

\begin{figure}[ht]
\begin{center}
\includegraphics*[height=4cm]{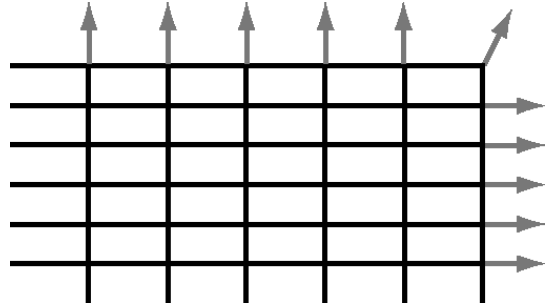}
\caption{The energy estimate for the semi-discrete initial-boundary
value problem on domains with corners shows that, in order to control
the growth due to the boundary term, boundary data must be given to
the incoming modes with respect to the unit normal $\vec n$.  At the
corner, the unit normal depends on the mesh spacings $h_1$ and $h_2$.}
\label{Fig:corner_normal}
\end{center}
\end{figure}

According to the discrete energy estimate above, to control the energy
growth due to the boundary term, one should give data to the incoming
variables in the direction $\vec n$, orthogonal to the boundary in
maximally dissipative form, as shown in Fig.~\ref{Fig:corner_normal}.
To define the unit normal at the corner of the grid we examine the
contribution to the energy estimate due to the corner point 
itself~\cite{Ols}.
We see that, for example, at $(i,j) = (N_1,N_2)$ we have
\begin{eqnarray}
&&\frac{h_2}{2} u^T_{N_1N_2} (HA^1u)_{N_1N_2} + \frac{h_1}{2} u^T_{N_1N_2}
(HA^2u)_{N_1N_2}  \nonumber\\
&&= \frac{|h|}{2}u^T_{N_1N_2}(HA^nu)_{N_1N_2},
\end{eqnarray}
where $|h| = \sqrt{h_1^2+h_2^2}$ and $\vec n = (h_2,h_1)/|h|$ is the
unit normal at $(N_1,N_2)$.  Similar results hold at the other
corners.  In particular, this shows that for uniform grids ($h_1=h_2$), data
should be given in the $45^\circ$ direction.

\subsubsection{Olsson's boundary conditions}

Let's assume that at the boundary there is one incoming, one outgoing,
and one zero speed mode and that $\Lambda = {\rm
diag}(+\lambda_+,-\lambda_-,0)$ with $\lambda_{\pm}>0$.  At each grid
point belonging to the boundary, boundary conditions are implemented
according to Olsson's prescription \cite{Ols}.  Namely, if $\vec
n=(n_1,n_2)$ is the outward pointing unit normal, we carry out the
following steps:
\begin{enumerate}
\item Compute $(W^{(+\lambda_+;n)}_{\rm old},W^{(-\lambda_-;n)}_{\rm
old},W^{(0;n)}_{\rm old})^T = Q(n)^T \Pi$, where $\Pi$ is the
discretized right hand side and $Q(n)$ is the orthogonal matrix that
diagonalizes the boundary matrix $HA^n$, $Q(n)^THA^nQ(n) = \Lambda$.
\item If the boundary condition at the continuum is $w^{(+\lambda_+;n)} =
Sw^{(-\lambda_-;n)} + g$, overwrite the ingoing and outgoing modes
according to
\begin{eqnarray*}
W^{(+\lambda_+;n)}_{\rm new} &=& \frac{S}{1+S^2}(S W^{(+\lambda_+;n)}_{\rm old} +
W^{(-\lambda_-;n)}_{\rm old}) + \frac{1}{1+S^2}\partial_t g\\
W^{(-\lambda_-;n)}_{\rm new} &=& \frac{1}{1+S^2}(S W^{(+\lambda_+;n)}_{\rm old} +
W^{(-\lambda_-;n)}_{\rm old}) - \frac{S}{1+S^2}\partial_t g
\end{eqnarray*}
and leave the zero speed mode unchanged, $W^{(0;n)}_{\rm new} =
W^{(0;n)}_{\rm old}$.  This will ensure that $W^{(+\lambda_+;n)}_{\rm new} = S
W^{(-\lambda_-;n)}_{\rm new} + \partial_t g$ and that the following linear
combination of in- and outgoing modes remains unchanged, $S
W^{(+\lambda_+;n)}_{\rm new} + W^{(-\lambda_-;n)}_{\rm new} = S W^{(+\lambda_+;n)}_{\rm old} +
W^{(-\lambda_-;n)}_{\rm old}$.  Note that unless $S=0$, the outgoing mode will
be modified.  When the exact solution is known, the boundary data
required to reproduce it are 
$g=g^{(+\lambda_+,n)}-Sg^{(-\lambda_-;n)}$, where $g^{(+\lambda_+;n)}$ and $g^{(-\lambda_-;n)}$ are
ingoing and outgoing characteristic variables of the exact solution.
\item The new modified rhs is obtained by multiplying
$(W^{(+\lambda_+;n)}_{\rm new},W^{(-\lambda_-;n)}_{\rm new},W^{(0;n)}_{\rm new})^T$
by $Q(n)$.
\end{enumerate}

\subsubsection{Consistency at corners}

Although giving data to the incoming variables at the corner in the
direction $\vec n$ controls the energy growth and therefore ensures
numerical stability, to achieve consistency with the boundary
conditions used at the two adjacent sides some extra care is required.
Let us assume that the normals to the two sides defining the corner
are $\vec n$ and $\vec m$ and that $\Lambda = {\rm diag}
(+\lambda_+,-\lambda_-,0)$ with $\lambda_{\pm} > 0$, i.e., on each side
there is one ingoing, one outgoing and one zero speed mode.  We give
data to the incoming variables at the sides according to
\begin{eqnarray}
w^{(+\lambda_+;n)}_{\rm new} &=& g^{(n)},
\label{Eq:nside}\\
w^{(+\lambda_+;m)}_{\rm new} &=& g^{(m)},
\label{Eq:mside}
\end{eqnarray}
where, for simplicity, we have assumed no coupling to the outgoing
fields.   At the continuum these two conditions will be
satisfied also at the corner.  Let us assume that at the corner data
is given in the direction $\vec p$.  We must translate
(\ref{Eq:nside}) and (\ref{Eq:mside}) in terms of characteristic
variables in the direction $\vec p$.  If $Q(r)$ denotes the orthogonal
matrix defining the characteristic variables in the generic direction
$\vec r$, $w^{(r)} = Q^T(r)u$, then we find that at the corner we must
use $w^{(\lambda_+;p)} = S w^{(-\lambda_-;p)} + g^{(p)}$, with a
non-trivial coupling
\begin{equation}
S = -\frac{[Q^T(m)Q(p)]_{13} [Q^T(n) Q(p)]_{12} - (n\leftrightarrow
m)}{[Q^T(m)Q(p)]_{13} [Q^T(n) Q(p)]_{11} - (n\leftrightarrow m) },
\label{Eq:Scorner}
\end{equation}
and boundary data
\begin{equation}
g^{(p)} = \frac{[Q^T(m)Q(p)]_{13} g^{(n)} - (n\leftrightarrow
m)}{[Q^T(m)Q(p)]_{13} [Q^T(n) Q(p)]_{11} - (n\leftrightarrow m) },
\end{equation}
where $[Q]_{ij}$ denotes the $ij$ matrix element of $Q$.  The notation
$(n\leftrightarrow m)$ indicates that the preceding term is repeated
with the exchange of the vectors $m$ and $n$.
In particular, if $g^{(n)}$ and $g^{(m)}$ vanish, then $g^{(p)}$ also
vanishes.  However, in general, the absence of coupling on the two
adjacent sides is not consistent with a vanishing $S$ at the corner,
Eq.~(\ref{Eq:Scorner}).

\subsection{The massless scalar field on a curved background}

\subsubsection{The axially symmetric system}

We now turn to the massless scalar field propagating on a curved
background $(M,g)$.  The equation of motion is the second order wave
equation
\begin{equation}
\nabla_{\mu} \nabla^{\mu} \Phi = 0,
\end{equation}
where $\nabla$ denotes the covariant derivative associated with the
Lorentz metric $g$.  In terms of the tensor density $\gamma^{\mu\nu} =
\sqrt{-g}g^{\mu\nu}$, the wave equation can be written
\begin{equation}
\partial_{\mu} (\gamma^{\mu\nu} \partial_{\nu} \Phi) = 0\,.
\label{Eq:wave2}
\end{equation}
We introduce the auxiliary variables $T= \partial_t \Phi$ and $d_i =
\partial_i \Phi$, and rewrite Eq.~(\ref{Eq:wave2}) in first order form,
\begin{eqnarray}
\partial_t \Phi &=& T\,,
\label{Eq:WEgen1}\\
\partial_t T &=& -\left(\gamma^{ti}\partial_i T +
\partial_i(\gamma^{it}T)+ \partial_i (\gamma^{ij}d_j) +\right.
\label{Eq:WEgen2}\\
&&+\left.\partial_t \gamma^{tt}T + \partial_t
\gamma^{tj}d_j\right)/\gamma^{tt},\nonumber\\
\partial_t d_i &=& \partial_i T\,.
\label{Eq:WEgen3}
\end{eqnarray}
The $\Phi$ component of a sufficiently smooth solution $(\Phi,T,d_i)$
of the first order system satisfies the second order wave equation
provided that the constraints $C_i \equiv d_i - \partial_i \Phi = 0$
are satisfied. An attractive feature of this particular first order
formulation is that the constraint variables propagate trivially, 
namely $\partial_t C_i = 0$~\cite{CalLehNeiPulReuSarTig}.  
In particular, this ensures that any
solution of (\ref{Eq:WEgen1}--\ref{Eq:WEgen3}) which satisfies the
constraints initially, will satisfy them at later times, even in the
presence of boundaries.

Since $\Phi$ does not appear in Eqs.~(\ref{Eq:WEgen2}--\ref{Eq:WEgen3}), 
we will drop Eq.~(\ref{Eq:WEgen1}) from the system.
The constraints are replaced by $C_{ij} \equiv \partial_{[i} d_{j]} =
0$, which also propagate trivially.  Interestingly, if
Eq.~(\ref{Eq:WEgen3}) and the constraints are discretized using
difference operators satisfying $[D_i,D_j] = 0$, which is usually the
case, then the time derivative of the discrete constraint variable
$C_{ij} = D_{[i}d_{j]}$ will also vanish.  In particular, for initial
data such that $d_i=0$, the discrete constraints will be identically
satisfied during evolution.

To simplify the problem we assume that the
background metric is axisymmetric, which implies that there exists a
spacelike Killing field $\mbox{\boldmath $\psi$} = \psi^{\mu}
\partial_{\mu}= \partial_{\phi}$.  
We always use coordinate systems adapted to the
Killing field, so that the
metric components are independent of the $\phi$ coordinate and, in
particular, $\partial_{\phi} \gamma^{\mu\nu} = 0$.  Since we are only
interested in axisymmetric solutions of the wave equation,
i.e., solutions which do not depend on $\phi$, the variable $d_{\phi}$
can be eliminated from the system.  Thus, the first order axisymmetric wave
equation consists of Eq.~(\ref{Eq:WEgen2}) and
(\ref{Eq:WEgen3}), where the Latin indices now span only two
dimensions, and one constraint.

\subsubsection{Characteristic speeds}

The characteristic speeds in an arbitrary direction $\vec n$, with
$|\vec n| = 1$, are given by the eigenvalues of
\begin{equation}
A^n = A^i n_i = \left(
\begin{array}{cc}
-2\gamma^{tn}/\gamma^{tt} & -
 \gamma^{nj}/\gamma^{tt}\\
n_i & 0
\end{array}
\right).
\end{equation}
These eigenvalues are $s_\pm = (\gamma^{tn} \pm \sqrt{(\gamma^{tn})^2
  - \gamma^{tt}\gamma^{nn}})/(-\gamma^{tt}) = \beta^n \pm \alpha
\sqrt{h^{nn}}$ and $s_0 = 0$, where $\alpha$ is the lapse function, 
$\beta^i$ the shift vector, and $h_{ij}$ is the induced 3-metric on
the $t=\mbox{const.}$ slices in the ADM decomposition~\cite{ADM}.
For the system to be hyperbolic it is
essential that $(\gamma^{tn})^2 - \gamma^{tt}\gamma^{nn} = h^{nn} \ge
0$, which will be true as long as the $t=\mbox{const.}$ hypersurfaces
are spacelike.  We also need $s_{\pm}$ to be bounded in the domain of
interest, which will be the case in a cylindrical or spherical
coordinate system (for $r\ge r_0 >0$), provided that the solution does
not depend on the azimuthal coordinate $\phi$.

\subsubsection{Symmetrizer, conserved energy, and characteristic variables}

One can verify that
\begin{equation}
H(t,\vec{x}) = \eta(t,\vec{x}) \left(
\begin{array}{cc}
-\gamma^{tt} & 0 \\
0 & \gamma^{ij}
\end{array}
\right)
\end{equation}
is the most general symmetric matrix that satisfies $HA^i = (HA^i)^T$.
When positive definite, which will be the case if and only if
$\partial_t$ is timelike and $\eta>0$, it represents the most general
symmetrizer of system (\ref{Eq:WEgen2}--\ref{Eq:WEgen3}).  If we use a
coordinate system adapted to the timelike Killing field $k =
\partial_t$, the components of $\gamma^{\mu\nu}$ will be time
independent.  In this case the symmetrizer
\begin{equation}
H= \left(
\begin{array}{cc}
-\gamma^{tt} & 0 \\
0 & \gamma^{ij} 
\end{array}
\right)
\label{Eq:Hsimple}
\end{equation}
satisfies Eq.~(\ref{Eq:conserved}) and gives rise to a conserved energy.

The boosting of the black hole will be performed by a Lorentz
transformation.  The time coordinate of the boosted frame will no
longer be adapted to the timelike Killing field.  In this case the
time derivative of
\begin{equation}
E = \int_{\Omega} (-\gamma^{tt}T^2 +
\gamma^{ij}d_id_j) \, d^2 x
\label{Eq:simpleE}
\end{equation}
is given by
\begin{eqnarray}
\frac{d}{dt} E &=& 2\int_{\partial\Omega} \left( T \gamma^{ti} T + T
\gamma^{ij} d_j \right) n_i \, d \sigma \\
&&+ \int_{\Omega} \left(T
\partial_t \gamma^{tt} T + 2T\partial_t \gamma^{tj} d_j + d_i
\partial_t \gamma^{ij} d_j \right) d^2x\nonumber
\end{eqnarray}
We assume that in a neighborhood of the outer boundary $\partial_t$ is
timelike.  The integrand of the surface term can be written as
\begin{equation}
 2( T \gamma^{ti} T + T \gamma^{ij} d_j) n_i = \lambda_+
{w^{(+\lambda_+;n)}}^2 - \lambda_- {w^{(-\lambda_-;n)}}^2
\end{equation}
where $\lambda_{\pm} = \gamma^n \pm \gamma^{tn}$ and 
\begin{eqnarray}
w^{(\pm \lambda_\pm; n)} &=& \pm \frac{\sqrt{1 \pm
\hat{\gamma}^{tn}}}{\sqrt{2}} T + \frac{1}{\sqrt{2}} \frac{
\hat{\gamma}^{in}d_i}{\sqrt{1 \pm \hat{\gamma}^{tn}}}\
\label{Eq:charvarspm}\\
w^{(0;n)} &=& \gamma_{\perp}^{i}d_i
\label{Eq:charvars0}
\end{eqnarray}
are the orthonormal characteristic variables of $HA^n$.  To simplify 
the notation we have introduced the quantities $\gamma^n =
\sqrt{\delta_{\mu\nu} \gamma^{\mu n}\gamma^{\nu n}}$,
$\hat{\gamma}^{\mu n} = \gamma^{\mu n}/\gamma^n$ and
$\gamma^i_{\perp}$.  The latter satisfies $\delta_{ij} \gamma^i_\perp
\gamma^j_\perp = 1$ and $\delta_{ij}\gamma^i_\perp \gamma^{jn} = 0$.
To express the primitive variables in terms of the characteristic
variables we invert Eqs.~(\ref{Eq:charvarspm})--(\ref{Eq:charvars0}),
\begin{eqnarray}
T &=& \frac{\sqrt{1+\hat{\gamma}^{tn}}}{\sqrt{2}} {w^{(+\lambda_+;n)}}
- \frac{\sqrt{ 1 -\hat{\gamma}^{tn}}}{\sqrt{2}} {w^{(-\lambda_-;n)}}
\label{Eq:charvarsT}
 \\ 
d_i &=& \frac{\hat{\gamma}^{in}}{\sqrt{2}} \left(
\frac{{w^{(+\lambda_+;n)}}}{\sqrt{1+\hat{\gamma}^{tn}}} +
\frac{{w^{(-\lambda_-;n)}}}{\sqrt{1-\hat{\gamma}^{tn}}} \right) +
\gamma^i_\perp w^{(0;n)}
\label{Eq:charvarsd_i}
\end{eqnarray}
Equations (\ref{Eq:charvarspm})--(\ref{Eq:charvarsd_i}) will be used in the
boundary conditions.

\subsubsection{Discretization}

Even when there is no conserved energy, it may be
desirable to discretize the right hand side of (\ref{Eq:WEgen2}) and
(\ref{Eq:WEgen3}) in a manner that satisfies the optimal estimate 
(\ref{Eq:goodestimate}), such as (\ref{Eq:splitting}) or other
alternatives, where the symmetrizer $H$ is given by (\ref{Eq:Hsimple}).

The discretization of the wave equation according to
(\ref{Eq:splitting}) leads to 
\begin{eqnarray*}
\partial_t T &=& 
-\left(\gamma^{ti}D_i T +
D_i(\gamma^{it}T)+ \frac{1}{2}D_i (\gamma^{ij}d_j)+  \right.\\
&& \left.+\frac{1}{2} \gamma^{ij}D_i d_j +
\frac{1}{2} \partial_i \gamma^{ij}d_j +
\partial_t \gamma^{tt}T + \partial_t
\gamma^{ti}d_i\right)/\gamma^{tt},\nonumber
\\
\partial_t d_i &=& \frac{1}{2} D_i T + \frac{1}{2}
({}^3\gamma^{-1})_{ik} D_j (\gamma^{kj} T) - \frac{1}{2} ({}^3
\gamma^{-1})_{ik} \partial_j \gamma^{kj} T\,,
\end{eqnarray*}
where ${}^3 \gamma^{-1}$ denotes the inverse of $\gamma^{ij}$.

Alternatively, one can simply replace the partial derivative $\partial_i$
in Eq.~(\ref{Eq:WEgen2}) and (\ref{Eq:WEgen3}) with the finite
difference operator $D_i$ satisfying (\ref{Eq:sumbyparts1}) and
(\ref{Eq:sumbyparts2}) and obtain the semi-discrete system 
\begin{eqnarray}
\partial_t T &=& -\left(\gamma^{ti}D_i T +
D_i(\gamma^{it}T)+ D_i (\gamma^{ij}d_j) +\right.\label{Eq:DWEgen2}\\
&&\left.+\partial_t \gamma^{tt}T + \partial_t
\gamma^{tj}d_j\right)/\gamma^{tt},\nonumber\\
\partial_t d_i &=& D_i T\,.
\label{Eq:DWEgen3}
\end{eqnarray}
which also satisfies the estimate (\ref{Eq:goodestimate}).  It is this
discretization that will be used throughout this work, even in the
boosted black hole case ($\partial_t \gamma^{\mu\nu} \neq 0$), where
the energy (\ref{Eq:simpleE}) is not conserved.  We analyze the
discretization at the axis of symmetry in the next sections.

\section{Minkowski background}
\label{Sec:WE_flat}

The energy method for constructing stable finite difference schemes
has, until recently, received little attention in numerical
relativity.  Thus we first present the wave equation in axisymmetric
Minkowski space to demonstrate the method, before moving to the more
complicated black hole configurations.  In this section we give energy
preserving discretizations for cylindrical and spherical coordinates.
In particular, we will show how to discretize the system on the axis
of symmetry in an energy conserving way.  The next
section examines discretizations for a Schwarzschild black hole in
Kerr--Schild spherical coordinates.

\subsection{Cylindrical coordinates}

\subsubsection{The system}

In a Minkowski background in cylindrical coordinates, $\{t,\rho,z,\phi\}$,
the second order axisymmetric wave equation has the form
\begin{equation}
\partial^2_t \Phi = \frac{1}{\rho} \partial_\rho \left( \rho \partial_\rho
\Phi \right) + \partial^2_z \Phi \,.
\end{equation}
We consider the first order formulation
\begin{eqnarray}
\partial_t T &=& \frac{1}{\rho} \partial_\rho (\rho P) + \partial_z Z, \label{Eq:WEqCyl1_1}\\
\partial_t P &=& \partial_{\rho} T,\label{Eq:WEqCyl1_2}\\
\partial_t Z &=& \partial_z T,\label{Eq:WEqCyl1_3}
\end{eqnarray}
where $T = \partial_t \Phi$, $P = \partial_\rho \Phi$ and $Z =
\partial_z \Phi$ are functions of $(t,\rho,z) \in [0,T]\times
[0,\rho_{\rm max}]\times [z_{\rm min},z_{\rm max}]$.

\subsubsection{Regularity conditions at the axis $\rho = 0$}

Smoothness at the $\rho=0$ axis requires that the odd
$\rho$-derivatives of the scalar field vanish on the axis, namely
$\partial^{2n-1}_\rho \Phi(t,\rho,z)|_{\rho=0} = 0$ for $n = 1, 2,
\ldots$.  This implies that the following conditions for the auxiliary
variables $T$, $P$, and $Z$, have to hold during evolution
\begin{eqnarray}
&&P|_{\rho=0} = \partial^{2n}_\rho P|_{\rho=0} = 0 \quad \mbox{for
$n=1,2,\ldots$}\label{Eq:oddP}\\
&&\partial^{2n-1}_\rho T|_{\rho=0} = \partial^{2n-1}_\rho Z|_{\rho=0} = 0 \quad
\mbox{for $n = 1, 2, \ldots$} \label{Eq:evenTZ}\;. 
\end{eqnarray}
If the initial data satisfies (\ref{Eq:oddP}) and
(\ref{Eq:evenTZ}), and the prescription $P(t,0,z) = 0$ is used as a
boundary condition at $\rho=0$, then the above conditions will hold at
later times.

\subsubsection{The boundary conditions}

Since in this coordinate system $\partial_t$ is a Killing field, the
energy (\ref{Eq:simpleE}) is conserved.   The time derivative of
\begin{equation}
E = \int_{z_{\rm min}}^{z_{\rm max}}\int_0^{\rho_{\rm
max}} \left( T^2 + P^2 + Z^2\right) \rho d\rho dz
\label{Eq:energy_axi}
\end{equation}
gives only boundary terms which can be controlled by giving
appropriate boundary data
\begin{eqnarray}
\frac{d}{dt} E &=& 2\int_{z_{\rm min}}^{z_{\rm max}} T(t,\rho_{\rm max},z)
R(t,\rho_{\rm max},z)\rho_{\rm max} dz \label{Eq:EdotContCyl}\\
&&+ 2\int_0^{\rho_{\rm max}}
\left[T(t,\rho,z)Z(t,\rho,z)\right]_{z=z_{\rm min}}^{z=z_{\rm max}}
\rho d\rho\;. \nonumber
\end{eqnarray}

\subsubsection{Energy conserving discretization}

The discretization of the right hand side of Eq.~(\ref{Eq:WEqCyl1_1}) at the
$\rho=0$ axis deserves special attention.  As a consequence of the
regularity conditions we have that  
\begin{equation}
\lim_{\rho\to 0^+}\frac{1}{\rho}\partial_\rho (\rho P) = 2
\partial_\rho P|_{\rho=0} \,.
\end{equation}
and therefore no infinities appear on the right hand side.

This suggests considering the semi-discrete approximation~\cite{Sar}
\begin{eqnarray}
\partial_t T_{ij} &=& \left\{
\begin{array}{ll}
2D^{(\rho)}_+ {P}_{0j} + D^{(z)} Z_{0j}, & i=0 \\
\frac{1}{\rho_i} D^{(\rho)} (\rho P)_{ij} + D^{(z)}Z_{ij}, & i \ge 1
\end{array}
\right.
\label{Eq:SDWE_flat_cyl1}\\
\partial_t {P}_{ij} &=& D^{(\rho)} T_{ij}, \quad i \ge 1
\label{Eq:SDWE_flat_cyl2}\\
\partial_t Z_{ij} &=& D^{(z)} T_{ij}, \quad i \ge 0
\label{Eq:SDWE_flat_cyl3}\;,
\end{eqnarray}
where $\rho_i = i\Delta \rho$ and $z_j = z_{\rm min} + j\Delta z$,
with $N_\rho \Delta \rho = \rho_{\rm max}$ and $N_z\Delta z = z_{\rm
max}-z_{\rm min}$.  The difference operators $D^{(\rho)}$ and
$D^{(z)}$ are second order accurate centered difference operators
where their computation does not involve points which do not belong to
the grid, and are first order accurate one sided difference operators
otherwise.  The regularity condition, ${P}_{0j} = 0$ for
$j=0,\ldots,N_z$, is enforced for all $t$, and Eq.~(\ref{Eq:oddP}) ensures
that $D_+^{(\rho)}P_{0j}$ is, in fact, a second order approximation.  A
solution of (\ref{Eq:SDWE_flat_cyl1}, \ref{Eq:SDWE_flat_cyl2},
\ref{Eq:SDWE_flat_cyl3}) conserves the discrete energy
\begin{eqnarray}
E &=& \sum_{j=0}^{N_z} \left[\sum_{i=1}^{N_\rho} \left(T_{ij}^2 +
{P}_{ij}^2 + Z_{ij}^2\right) \rho_i \sigma_i \Delta \rho \right.
\label{Eq:discEcyl}\\
&&\left.+ \frac{1}{4}
\left(T_{0j}^2 + Z_{0j}^2\right) \Delta \rho^2 \right] \sigma_j \Delta z\,,\nonumber
\end{eqnarray}
which is consistent with the continuum expression
(\ref{Eq:energy_axi}).  More precisely, using the fact that $\partial_t
T_{0j} = \frac{2}{\Delta \rho}{P}_{1j} + D^{(z)}Z_{0j}$ and the basic
properties of the finite difference operators, one can see
that the following estimate
\begin{eqnarray}
\frac{d}{dt}E &=& 2\sum_{j=0}^{N_z} T_{N_\rho j}\rho_{N_\rho} {P}_{N_\rho
j} \sigma_j \Delta z \label{Eq:CSE}\\
&&+ 2\sum_{i=1}^{N_\rho}
(T_{iN_z}Z_{iN_z}-T_{i0}Z_{i0})\rho_i \sigma_i \Delta \rho \nonumber\\
&&+\frac{1}{2}(T_{0N_z}Z_{0N_z}-T_{00}Z_{00}) \Delta \rho^2\nonumber
\end{eqnarray}
holds, consistently with the continuum limit (\ref{Eq:EdotContCyl}).

As it is pointed out in Section 12.7 of \cite{GKO-Book}, one
order less accuracy at the boundary is allowed, in the sense that it
does not affect the overall accuracy of the scheme, provided that the
physical boundary conditions are approximated to the same order as the
differential operators at the inner points.

\subsubsection{Discrete boundary conditions}
\begin{figure}[ht]
\begin{center}
\includegraphics*[height=8cm]{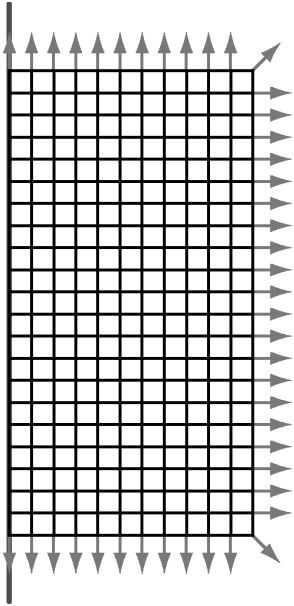}
\caption{This figure shows how the unit normal at the boundary grid
points should be chosen.  We note that at the corners which lie on the
axis of symmetry we must apply both the regularity and the boundary
conditions.}
\label{Fig:cyl_grid}
\end{center}
\end{figure}

By inspecting the boundary terms of the discrete energy estimate
(\ref{Eq:CSE}), we can readily see how the boundary data should be
given at each boundary grid point ($j=0$, $j=N_z$, and $i=N_{\rho}$).
In the case of a uniform grid ($\Delta \rho = \Delta z$), in order to
control the energy growth boundary data should be given in maximally
dissipative form in the directions shown in Fig.~\ref{Fig:cyl_grid}.

The presence of lower order terms in
(\ref{Eq:discEcyl}), in addition to ensuring that the discrete energy
is positive definite on the axis, indicates how to specify boundary
data at the corner grid points that lie on the axis.

\subsection{Spherical coordinates}

In this section we discretize the wave equation in Minkowski space
with spherical coordinates $\{t,r,\theta, \phi\}$.

\subsubsection{The system}

The second order axisymmetric wave equation on a flat background in
spherical coordinates 
\begin{equation}
\partial_t^2 \Phi = \frac{1}{r^2} \partial_r(r^2 \partial_r \Phi) +
\frac{1}{r^2 \sin\theta} \partial_{\theta} ( \sin\theta
\partial_{\theta} \Phi),
\end{equation}
is written in first order form as
\begin{eqnarray}
\partial_t T &=& \frac{1}{r^2} \partial_r (r^2R) +
\frac{1}{r^2\sin\theta} \partial_{\theta} (\sin\theta \Theta)
\label{Eq:WEqSph1_1}\\
\partial_t R &=& \partial_r T 
\label{Eq:WEqSph1_2}\\
\partial_t \Theta &=& \partial_{\theta} T,
\label{Eq:WEqSph1_3}
\end{eqnarray}
where $T = \partial_t \Phi$, $R = \partial_r \Phi$ and $\Theta =
\partial_{\theta} \Phi$ are functions of $(t,r,\theta) \in [0,T]\times
[r_{\rm min},r_{\rm max}] \times [0,\pi]$.

\subsubsection{Regularity conditions on the axis $\theta = 0$ and $\theta=\pi$}

Smoothness requires that the odd $\theta$-derivatives of the scalar
field vanish on the axis of symmetry, namely at $\theta = 0$ and
$\theta = \pi$.  This implies that
\begin{eqnarray}
&&\Theta|_{\theta = 0,\pi} = \partial_\theta^{2n} \Theta|_{\theta = 0,
\pi} = 0 \qquad n=1, 2, \ldots \label{Eq:oddTheta}\\
&&\partial_\theta^{2n-1} T|_{\theta = 0,\pi} = \partial_\theta^{2n-1}
R|_{\theta = 0,\pi} = 0 \qquad n = 1, 2, \ldots\label{Eq:evenTR}
\end{eqnarray}
As in the cylindrical case, it is possible to show that if the initial
data satisfy (\ref{Eq:oddTheta}) and (\ref{Eq:evenTR}) and the
boundary conditions $\Theta|_{\theta = 0,\pi} = 0$ are used during
evolution, then the above regularity conditions will continue to hold.

\subsubsection{Boundary conditions}

Since we are interested in a domain of the form $\Omega = \{
(r,\theta) \in \mathbb{R}^2 | r_{\rm min}\le r \le r_{\rm max}, 0 \le \theta
\le \pi \}$, where $r_{\rm min} > 0$, the characteristic speeds are
bounded by $\max\{ 1, r^{-1}_{\rm min}\}$.  The conserved energy is
\begin{equation}
E = \int_{r_{\rm min}}^{r_{\rm max}} \int_0^{\pi} \left( T^2 + R^2 +
\frac{\Theta^2}{r^2} \right) r^2 \sin\theta d\theta dr,
\end{equation}
and its time derivative is given by
\begin{equation}
\frac{d}{dt} E  = 2 \int_0^{\pi} \left. r^2 RT \right|_{r=r_{\rm min}}^{r = r_{\rm max}}
\sin\theta d\theta .
\label{Eq:EdotspherM}
\end{equation}
At $r=r_{\rm max}$ and $r=r_{\rm min}$ boundary data must be given to
the incoming modes.

\subsubsection{Energy conserving discretization}

As a consequence of the regularity conditions on the axis of symmetry,
we have that 
\begin{equation}
\lim_{\theta \to m\pi} \frac{1}{\sin\theta} \partial_\theta
(\sin\theta \Theta) = 2 \partial_\theta \Theta|_{\theta = m\pi}, \qquad
m = 0, 1.
\end{equation}
We discretize the right hand side of 
(\ref{Eq:WEqSph1_1}--\ref{Eq:WEqSph1_3}) as
\begin{widetext}
\begin{eqnarray}
\partial_t T_{ij}&=& \left\{
\begin{array}{ll}
\frac{1}{r_i^2} D^{(r)}(r^2R)_{ij} +
\frac{2}{r_i^2}D^{(\theta)}_{\pm} \Theta_{ij} & j = 0, N_\theta\\
\frac{1}{r_i^2} D^{(r)}(r^2R)_{ij} + \frac{1}{r_i^2 \sin\theta_j}
D^{(\theta)}_0(\sin\theta \Theta)_{ij} & j=1,\ldots, N_\theta-1 
\end{array}\right.\\
\partial_t R_{ij} &=& D^{(r)} T_{ij},\\
\partial_t \Theta_{ij} &=& D^{(\theta)} T_{ij},
\end{eqnarray}
\end{widetext}
where $r_i = r_{\rm min} + i \Delta r$ and $\theta_j = j\Delta
\theta$, with $N_r \Delta r = r_{\rm max} - r_{\rm min}$ and
$N_{\theta} \Delta\theta = \pi$.  The condition $\Theta_{i0} =
\Theta_{iN_{\theta}} = 0$ on the axis is enforced at all times.  The
following discrete energy
\begin{eqnarray}
E &=& \sum_{i=0}^{N_r} \sum_{j=1}^{N_\theta-1} \left( T^2_{ij} +
R^2_{ij} + \frac{\Theta^2_{ij}}{r_i^2} \right) r_i^2
\sin\theta_j \sigma_i \Delta \theta \Delta r \\
&& + \frac{1}{2}
\sum_{i=0}^{N_r} \left( T^2_{i0} + R^2_{i0}\right)r_i^2 \sigma_i
\sin\Delta\theta \frac{\Delta\theta}{2} \Delta r \nonumber \\
&& + \frac{1}{2}
\sum_{i=0}^{N_r} \left( T^2_{iN_\theta} + R^2_{iN_\theta}\right)r_i^2
\sigma_i \sin\Delta\theta \frac{\Delta\theta}{2} \Delta r, \nonumber
\end{eqnarray}
is conserved by the semi-discrete system.  Its time derivative
\begin{eqnarray}
\frac{d}{dt} E &=& 2\sum_{j=1}^{N_\theta-1} \left.\left( T_{ij} R_{ij} r^2_i
\right)\right|_{i=0}^{i=N_r} \sin\theta_j \Delta\theta \\
&& + \left.\left( T_{i0} R_{i0} r^2_i
\right)\right|_{i=0}^{i=N_r} \sin\Delta\theta  \frac{\Delta\theta}{2} \nonumber\\
&& + \left.\left( T_{iN_\theta} R_{iN_\theta} r^2_i
\right)\right|_{i=0}^{i=N_r} \sin\Delta\theta
\frac{\Delta\theta}{2}, \nonumber
\end{eqnarray}
 gives only boundary terms consistently with the continuum estimate
(\ref{Eq:EdotspherM}).

\subsubsection{Discrete boundary conditions}
\begin{figure}[ht]
\begin{center}
\includegraphics*[height=8cm]{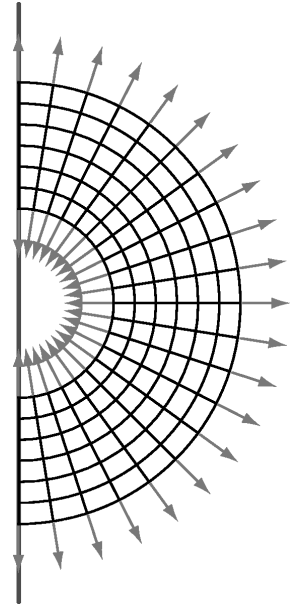}
\caption{According to the discrete energy estimate, boundary data should
be given to the incoming variable in the direction indicated in the
figure.}
\label{Fig:spher_grid}
\end{center}
\end{figure}

The choice of unit normal at the boundary grid points ($i=0$ and
$i=N_r$) is illustrated in Fig.~\ref{Fig:spher_grid}.

\section{Fixed black hole background}
\label{Sec:WE_BH}
This section is a generalization of the results of the previous
section to the case of a static black hole background.  The background
metric is Schwarzschild in Kerr--Schild coordinates~\cite{KerSch}.
The Cartesian components of the background metric can be written as 
\begin{equation}
g_{\mu\nu} = \eta_{\mu\nu} + \frac{2M}{r}\ell_{\mu}\ell_{\nu},
\label{Eq:KSmetric}
\end{equation}
where $\eta_{\mu\nu} = {\rm diag} (-1,+1,+1,+1)$, $r^2 = x^2 + y^2 +z^2$ and
$\ell_{\mu} = (1,\vec{x}/r)$.   In these coordinates the determinant
of the four-metric is $g=-1$.

The tensor density components $\gamma^{\mu\nu}$, which are needed to
write down the 3D wave equation in first order form, are given by
\begin{equation}
\gamma^{\mu\nu} = \eta^{\mu\nu}
-\frac{2M}{r}\ell^{\mu}\ell^{\nu},
\end{equation}
where $\ell^{\mu} = \eta^{\mu\nu} \ell_{\nu} = (-1,\vec{x}/r)$.  

As we do not wish to consider cylindrical excision in this paper,
we analyze here only the spherical coordinate case.

\subsection{Spherical coordinates}

\subsubsection{The system}

In spherical Kerr--Schild coordinates the components of $\gamma^{\mu\nu}$ on a
Schwarzschild background are
\begin{eqnarray}
\gamma^{\mu\nu} &=& r^2\sin\theta
\left(\eta^{\mu\nu} - \frac{2M}{r} l^{\mu} l^{\nu} \right)\,, \\
\eta^{\mu\nu} &=& {\rm diag}\left\{ -1, +1, +\frac{1}{r^2},
+\frac{1}{r^2\sin^2\theta}\right\}\,,\nonumber\\
\ell^{\mu} &=& (-1,+1,0,0).\nonumber
\end{eqnarray}
The first order axisymmetric wave equation is
\begin{eqnarray}
\partial_t T &=& \frac{2M}{r^+}\partial_rT +
\frac{2M}{rr^+}\partial_r(rT) + \frac{1}{rr^+}
\partial_r(rr^-R) \label{Eq:WE_BH_spher1}\\
&&+\frac{1}{rr^+\sin\theta}\partial_{\theta}(\sin\theta \Theta)
\nonumber\\ 
\partial_t R &=& \partial_r T
\label{Eq:WE_BH_spher2}\\
\partial_t \Theta &=& \partial_{\theta} T,
\label{Eq:WE_BH_spher3}
\end{eqnarray}
where $r^{\pm} = r\pm 2M$, $T = \partial_t \Phi$, $R = \partial_r
\Phi$ and $\Theta = \partial_{\theta} \Phi$.  In the region of
interest, $\Omega = \{ (r,\theta) \in \mathbb{R}^2 | 2M \le r \le
r_{\rm max}, 0 \le \theta \le \pi \}$, with $M>0$, the characteristic
speeds are bounded.

\subsubsection{Regularity conditions on the axis $\theta = 0$ and $\theta=\pi$}

Smoothness requires that the odd $\theta$-derivatives of the scalar
field vanish on the axis of symmetry.  This implies that
\begin{eqnarray}
&&\Theta|_{\theta = 0,\pi} = \partial_\theta^{2n} \Theta|_{\theta = 0,
\pi} = 0 \qquad n=1, 2, \ldots \label{Eq:oddTheta_BH}\\
&&\partial_\theta^{2n-1} T|_{\theta = 0,\pi} = \partial_\theta^{2n-1}
R|_{\theta = 0,\pi} = 0 \qquad n = 1, 2, \ldots\label{Eq:evenTR_BH}
\end{eqnarray}

\subsubsection{The boundary conditions}
A symmetrizer which gives rise to a conserved energy is given by
\begin{equation}
H = {\rm diag} \{ rr^+\sin\theta, rr^-\sin\theta , \sin\theta \},
\end{equation}
which is positive definite for $0<\theta<\pi$ and $r>2M$.  Inside the
event horizon, $r<2M$, the vector field $\partial_t$ becomes spacelike
and the system in only strongly hyperbolic.

The time derivative of the energy
\begin{equation}
E = \int_{2M}^{r_{\rm max}} \int_0^{\pi} \left(
rr^+T^2 + rr^-R^2 + \Theta^2 \right) \sin\theta d\theta dr,
\end{equation}
gives only boundary terms
\begin{equation}
\frac{d}{dt} E  = 2\int_0^{\pi} \left.\left( 2MrT^2 + rr^- TR \right)\right|_{r=2M}^{r = r_{\rm max}}
\sin\theta d\theta
\end{equation}

In addition to the regularity condition $\Theta = 0$ at the axis, the
problem requires boundary data at $r=r_{\rm max}$. 

\subsubsection{Energy conserving discretization}

We discretize the right hand side of 
(\ref{Eq:WE_BH_spher1}--\ref{Eq:WE_BH_spher3}) as
\begin{widetext}
\begin{eqnarray}
\partial_t T_{ij}&=& \left\{
\begin{array}{ll}
\frac{2M}{r^+_i}D^{(r)}T_{ij} + \frac{2M}{r_ir^+_i} D^{(r)}(rT)_{ij} +
\frac{1}{r_ir^+_i} D^{(r)}(rr^-R)_{ij} +
\frac{2}{r_ir^+_i}D^{(\theta)}_{\pm} \Theta_{ij} & j = 0, N_\theta\\
\frac{2M}{r^+_i}D^{(r)}T_{ij} + \frac{2M}{r_ir^+_i} D^{(r)}(rT)_{ij} +
\frac{1}{r_ir^+_i} D^{(r)}(rr^-R)_{ij} +
\frac{1}{r_ir^+_i \sin\theta_j}
D^{(\theta)}_0(\sin\theta \Theta)_{ij} & j=1,\ldots, N_\theta-1 
\end{array}\right.\\
\partial_t R_{ij} &=& D^{(r)} T_{ij}\\
\partial_t \Theta_{ij} &=& D^{(\theta)} T_{ij},
\end{eqnarray}
\end{widetext}
where $r_i = 2M + i \Delta r$ and $\theta_j = j\Delta
\theta$, with $N_r \Delta r = r_{\rm max} - 2M$ and
$N_{\theta} \Delta\theta = \pi$.  The following discrete energy,
\begin{eqnarray*}
E &=& \sum_{i=0}^{N_r} \sum_{j=1}^{N_\theta-1} \left( r_ir^+_iT^2_{ij}
+ r_ir^-_iR^2_{ij} + \Theta^2_{ij} \right) \sin\theta_j \sigma_i
\Delta \theta \Delta r\\
&& + \frac{1}{2} \sum_{i=0}^{N_r} \left( r_ir^+_iT^2_{i0} +
r_ir^-_iR^2_{i0}\right) \sigma_i
\sin\Delta\theta \frac{\Delta\theta}{2} \Delta r \nonumber \\
&& + \frac{1}{2} \sum_{i=0}^{N_r} \left( r_ir^+_iT^2_{iN_\theta} +
r_ir^-_iR^2_{iN_\theta}\right) \sigma_i
\sin\Delta\theta \frac{\Delta\theta}{2} \Delta r, \nonumber
\end{eqnarray*}
is conserved. Its time derivative is given by
\begin{eqnarray*}
\frac{d}{dt} E &=& 2\sum_{j=1}^{N_\theta-1} \left.\left( 2Mr_iT^2_{ij} + r_ir^-_iT_{ij} R_{ij} 
\right)\right|_{i=0}^{i=N_r} \sin\theta_j \Delta\theta \\
&& + \left.\left( 2Mr_iT^2_{i0} + r_ir^-_iT_{i0} R_{i0} 
\right)\right|_{i=0}^{i=N_r} \sin\Delta\theta  \frac{\Delta\theta}{2} \nonumber\\
&& + \left.\left( 2Mr_iT^2_{iN_{\theta}} + r_ir^-_iT_{iN_\theta} R_{iN_\theta}
\right)\right|_{i=0}^{i=N_r} \sin\Delta\theta
\frac{\Delta\theta}{2}. \nonumber
\end{eqnarray*}
We point out that, since $\Theta_{i0}= \Theta_{iN_{\theta}} = 0$, the
discrete energy is positive definite on the axis.  However, because
$r^-_0 = 0$, it does not control the growth of $R_{0j}$ on the event
horizon.  Numerical experiments indicate that this does not cause any
problems.  Moreover, experiments do not suggest that placing $r_{\rm min}$
within the horizon, where the equations are only strongly hyperbolic,
leads to an unstable scheme.

\subsubsection{Discrete boundary conditions}
Data should be given to the incoming modes at $r=r_{\rm max}$ as in
Fig.~\ref{Fig:spher_grid}.  Unlike the Minkowski case, no boundary
conditions should be given when the inner boundary, $r=r_{\rm min}$,
 is at or within the event horizon.

\section{Boosted black hole background}
\label{Sec:Boosted}

Finally, we consider the case in which the scalar field propagates on
a boosted black hole background.  To solve this problem
we introduce two coordinate patches:   one patch fixed
to the outer boundaries and one patch co-moving with the black hole,
and fixed to the inner excision boundary.  We choose cylindrical
coordinates for the first coordinate patch, boosted with respect
to the black hole such that the hole moves with velocity $\beta$
along the symmetry axis in these coordinates.  Spherical coordinates
are used on the second patch.  These coordinates are chosen by
fixing the event horizon at a constant coordinate value ($r'=2M$),
and requiring that all data in both coordinate systems are simultaneous.
By adapting these coordinates to the black hole horizon, we may 
excise the spherical grid at $r'=2M$ for all values of the boost parameter.
In this section we first write down the the components of the 4-metric 
in a boosted Cartesian coordinate system, then discuss the
two coordinate systems in some detail.  

We recall that in a Cartesian coordinate
system $\{ t, x, y, z\}$, with respect to which the black hole is at
rest, the metric components have the form given in (\ref{Eq:KSmetric}).
Under a Lorentz boost, i.e., in the new coordinates
\begin{eqnarray}
\bar{t} &=& \gamma(t - \beta z )\\
\bar{x} &=& x\nonumber\\
\bar{y} &=& y\nonumber\\
\bar{z} &=& \gamma (z - \beta t),\nonumber
\end{eqnarray}
where $\gamma = (1-\beta^2)^{-1/2}$, the components of the Kerr--Schild
metric become
\begin{eqnarray*}
g_{\bar\mu \bar\nu} &=& \eta_{\bar\mu \bar\nu} + \frac{2M}{r}
\ell_{\bar\mu}\ell_{\bar\nu}\,, \\
\qquad \eta_{\bar\mu\bar\nu} &=& {\rm diag} \{ -1, +1, +1, +1 \}\,,\\
\qquad \ell_{\bar\mu} &=& \left(
\hat r, \bar x, \bar y, \hat z\right)/r,
\end{eqnarray*}
where $\hat r = \gamma (r+\beta z)$, $\hat z = \gamma (z+ \beta r)$,
$z = \gamma (\bar{z} + \beta \bar{t})$ and $r^2 = x^2+y^2 +z^2 = 
\bar{x}^2 + \bar{y}^2 + \gamma^2 (\bar{z} + \beta \bar{t})^2$.  At
time $\bar{t}$ the singularity is located at
$(\bar{x},\bar{y},\bar{z}) = (0,0,-\beta \bar{t})$.

\subsection{Boosted cylindrical coordinates}

\begin{figure}[ht]
\begin{center}
\includegraphics*[height=10cm]{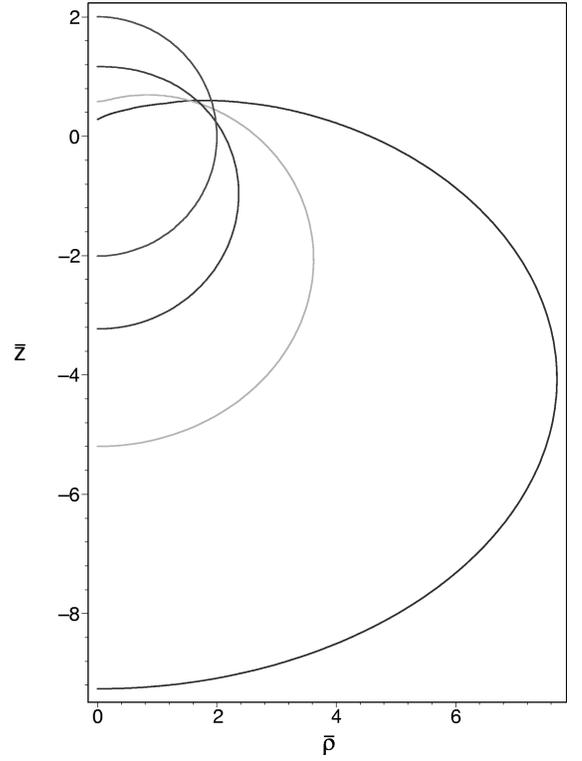}
\caption{The regions delimited by the curves are regions in which the
system in not symmetrizable hyperbolic, but only strongly hyperbolic
for $\beta=0, -1/4, -1/2, -3/4$ (the black hole is located at $(\bar
\rho, \bar z) = (0,0)$ and moves in the $+\bar z$ direction).  As the
boost parameter $\beta$ increases in magnitude there is a larger part
of the cylindrical domain in which the system in only strongly
hyperbolic.}
\label{Fig:SHRegions}
\end{center}
\end{figure}

We now choose cylindrical coordinates $\{ \bar{t}, \bar{\rho},
\bar{z}, \bar{\phi} \}$, with $\bar{\rho} \cos\bar{\phi} = \bar{x}$
and $\bar{\rho} \sin\bar{\phi} = \bar{y}$, giving
\begin{eqnarray*}
g_{\bar\mu \bar\nu} &=& \eta_{\bar\mu \bar\nu} + \frac{2M}{r}
\ell_{\bar\mu}\ell_{\bar\nu}\,, \\
\eta_{\bar\mu\bar\nu} &=& {\rm diag} \{ -1, +1, +1, +\bar\rho^2 \}\,,\\
\ell_{\bar\mu} &=& \left(
\hat r, \bar \rho, \hat z, 0 \right)/r
\end{eqnarray*}
and 
\begin{eqnarray}
\gamma^{\bar\mu \bar\nu} &=& \bar\rho \left(\eta^{\bar\mu\bar\nu} -
\frac{2M}{r}
\ell^{\bar\mu}\ell^{\bar\nu}\right)\,,\label{Eq:gamma_cyl_boosted}\\
\eta^{\bar\mu\bar\nu} &=& {\rm diag} \{ -1, +1,
+1, +\frac{1}{\bar\rho^2} \}\,,\nonumber\\
\ell^{\bar\mu} &=&
\left( -\hat r, \bar \rho, \hat z, 0 \right)/r,\nonumber
\end{eqnarray}
where $z = \gamma(\bar{z} + \beta\bar{t})$ and $r^2 = \bar{\rho}^2+
\gamma^2 (\bar{z} + \beta \bar{t})^2$.  Unfortunately, in these
coordinates the wave equation has a rather unpleasant form:  the
components of $\gamma^{\bar\mu\bar\nu}$ have a non-trivial dependence
on the three coordinates $\bar{\rho}$ and $\bar{z}$, and especially,
$\bar t$.

The analytic expressions for the time derivatives of the
$\gamma^{\bar{t}\bar{\mu}}$ components are needed.  Using the fact
that $\partial_{\bar{t}} r = \beta\gamma z/r$, $\partial_{\bar{t}}
\hat r = \beta\gamma \hat z/r$ and $\partial_{\bar{t}} \hat z =
\beta\gamma \hat r/r$ we get
\begin{eqnarray}
\partial_{\bar{t}} \gamma^{\bar{t}\bar{t}} &=& 
2M\beta\gamma \bar\rho \hat r (3z\hat r- 2r\hat z )/r^5,\\
\partial_{\bar{t}} \gamma^{\bar{t}\bar{\rho}} &=&
2M\beta\gamma \bar{\rho}^2 ( r\hat z - 3z\hat r)/r^5,\nonumber\\
\partial_{\bar{t}} \gamma^{\bar{t}\bar{z}} &=&
2M\beta\gamma\bar{\rho} \left( r\hat z^2 +
r\hat r^2 - 3z \hat r\hat z \right)/r^5.\nonumber
\end{eqnarray}

In this coordinate system our first order formulation has no conserved
energy ($\partial_{\bar{t}} \gamma^{\bar\mu\bar\nu} \neq 0$).  The
region in which the system is symmetrizable hyperbolic is determined
by the set of points in which $\partial_t$ is timelike,
\begin{equation}
-g_{\bar t \bar t} = 1-\frac{2M\hat r^2}{r^3} > 0.
\end{equation}
Fig.~\ref{Fig:SHRegions} shows the regions of lack of symmetric
hyperbolicity for different values of the boost parameter.

On the axis of symmetry ($\bar{\rho} = 0$) the equations need to be
expressed in a form which avoids ``$0/0$''.  This can be done by taking
the limit $\bar{\rho} \to 0$ in the equations.  It is convenient to
introduce the quantities
\begin{eqnarray*}
&&\tilde{\gamma}^{\bar{t}\bar{t}} =
\frac{\gamma^{\bar{t}\bar{t}}}{\bar{\rho}},\quad
\tilde{\gamma}^{\bar{t}\bar{\rho}} =
\frac{\gamma^{\bar{t}\bar{\rho}}}{\bar{\rho}^2},\quad
\tilde{\gamma}^{\bar{t}\bar{z}} =
\frac{\gamma^{\bar{t}\bar{z}}}{\bar{\rho}},\quad\\
&&\tilde{\gamma}^{\bar{\rho}\bar{\rho}} =
\frac{\gamma^{\bar{\rho}\bar{\rho}}}{\bar{\rho}},\quad
\tilde{\gamma}^{\bar{\rho}\bar{z}} =
\frac{\gamma^{\bar{\rho}\bar{z}}}{\bar{\rho}^2},\quad
\tilde{\gamma}^{\bar{z}\bar{z}} =
\frac{\gamma^{\bar{z}\bar{z}}}{\bar{\rho}},
\end{eqnarray*}
which have a finite limit for $\bar{\rho} \to 0$ (since the
singularity is excised we can assume that $r \ge r_0>0$).  The
right hand side of (\ref{Eq:WEgen2}) at $\bar\rho = 0$ becomes
\begin{eqnarray*}
\partial_{\bar{t}} \bar{T} &=& \left( \tilde{\gamma}^{\bar{t}\bar{z}}
\partial_{\bar{z}} \bar{T} + 2 \tilde{\gamma}^{\bar{t} \bar{\rho}}
\bar{T} + \partial_{\bar{z}} (\tilde{\gamma}^{\bar{t} \bar{z}}
\bar{T}) + 2 \tilde{\gamma}^{\bar{\rho} \bar{\rho}}
\partial_{\bar{\rho}} \bar{P} +\right.\\
&&\left.+ 2 \tilde{\gamma}^{\bar{\rho} \bar{z}}
\bar{Z} + \partial_{\bar{z}} (\tilde{\gamma}^{\bar{z} \bar{z}}
\bar{Z}) + \partial_{\bar{t}} \tilde{\gamma}^{\bar{t} \bar{t}} \bar{T}
+ \partial_{\bar{t}} \tilde{\gamma}^{\bar{t} \bar{z}} \bar{Z}\right)
/(-\tilde{\gamma}^{\bar{t} \bar{t}})\,.
\end{eqnarray*}

\subsection{Co-moving spherical coordinate system}

We introduce a spherical coordinate system $\{t',r',\theta',\phi'\}$
which is related to the unboosted Cartesian coordinates $\{t,x,y,z\}$,
the coordinates in which the black hole is at rest, by
\begin{eqnarray}
t' &=& \bar t = \gamma(t - \beta z)\label{Eq:sc1}\\
r' &=& \sqrt{x^2+y^2+z^2}\nonumber\\
\theta' &=& \cos^{-1} \left(\frac{z}{\sqrt{x^2+y^2+z^2}}\right)\nonumber\\
\phi' &=& \tan^{-1} \left(\frac{y}{x}\right).\nonumber
\end{eqnarray}
We emphasize that (\ref{Eq:sc1}) is not a Lorentz
transformation.  The coordinates are adapted to the event horizon in
the sense that its location ($r'=2M$) is time independent, and setting
$t'=\bar t$ maintains simultaneity in the two coordinate systems.
  The metric
components and the components of $\gamma^{\mu'\nu'}$ are more
conveniently written in matrix form
\begin{widetext}
\[
g_{\mu'\nu'} = \left(
\begin{array}{cccc}
\frac{1}{\gamma^2}\left(-1+\frac{2M}{r'}\right) & \frac{1}{\gamma}\left(
\frac{2M}{r'} - \beta \cos \theta' \left(1-\frac{2M}{r'}\right) \right) &
\frac{1}{\gamma}(r'-2M)\beta\sin\theta' & 0\\
\cdot & (1+\beta\cos\theta') \left(1-\beta\cos\theta' + \frac{2M}{r'} (1
+ \beta\cos\theta')\right) & \beta\sin\theta' \left(
(r'-2M)\beta\cos\theta' - 2M\right) & 0 \\
\cdot & \cdot & r'(r'-\beta^2\sin^2\theta'(r'-2M)) & 0 \\
\cdot & \cdot & \cdot & r'^2 \sin^2\theta'
\end{array}
\right)
\]
\begin{equation}
\gamma^{\mu'\nu'} = \left(
\begin{array}{cccc}
-\frac{1}{\gamma} r' \sin\theta'\left(r'+2 M \gamma^2
(1+\beta\cos\theta')^2\right) & r'\sin\theta' \left( 2M -
\beta\cos\theta' (r'-2M)\right) & \beta r' \sin^2\theta' & 0 \\
\cdot & \frac{1}{\gamma} r' (r'-2M) \sin\theta' & 0 & 0 \\
\cdot & \cdot & \frac{\sin\theta'}{\gamma} & 0 \\
\cdot & \cdot & \cdot & \frac{1}{\gamma\sin\theta'}
\end{array}
\right).
\label{Eq:gamma_spher_comoving}
\end{equation}
\end{widetext}
We have symmetric hyperbolicity for $r' > 2M$ and $0 < \theta' < \pi$.

On the axis of symmetry ($\theta'= 0$ or $\theta' = \pi$) the
equations need to be expressed in a form that avoids ``$0/0$''.  This
can be done by taking the limit $\theta' \to \theta_0$, where
$\theta_0 = 0, \pi$ in the equations.  If we introduce the quantities
\begin{eqnarray*}
&&\tilde{\gamma}^{t't'} =
\frac{\gamma^{t't'}}{\sin\theta'},\quad
\tilde{\gamma}^{t'r'} =
\frac{\gamma^{t'r'}}{\sin\theta'},\quad
\tilde{\gamma}^{t'\theta'} =
\frac{\gamma^{t'\theta'}}{\sin^2\theta'},\\
&&\tilde{\gamma}^{r'r'} =
\frac{\gamma^{r'r'}}{\sin\theta'},\quad
\tilde{\gamma}^{\theta'\theta'} =
\frac{\gamma^{\theta'\theta'}}{\sin\theta'},
\end{eqnarray*}
which have a finite limit for $\sin\theta' \to 0$, then the right hand
side of (\ref{Eq:WEgen2}) on axis becomes
\begin{eqnarray}
\partial_{t'} T' &=& \left( \tilde{\gamma}^{t'r'}
\partial_{r'} T' + \partial_{r'}(\tilde{\gamma}^{t'r'}T') \pm
2 \tilde{\gamma}^{t' \theta'}
T'  \right.\\
&&\left.+ \partial_{r'} (\tilde{\gamma}^{r'r'}
R')+ 2 \tilde{\gamma}^{\theta'\theta'}
\partial_{\theta'} \Theta' \right)
/(-\tilde{\gamma}^{t' t'}),\nonumber
\end{eqnarray}
where the components of $\tilde{\gamma}^{\mu'\nu'}$ are understood to be
evaluated at $\theta'=0, \pi$.

\subsection{Overlapping grids}

\begin{figure}[ht]
\begin{center}
\includegraphics*[height=10cm]{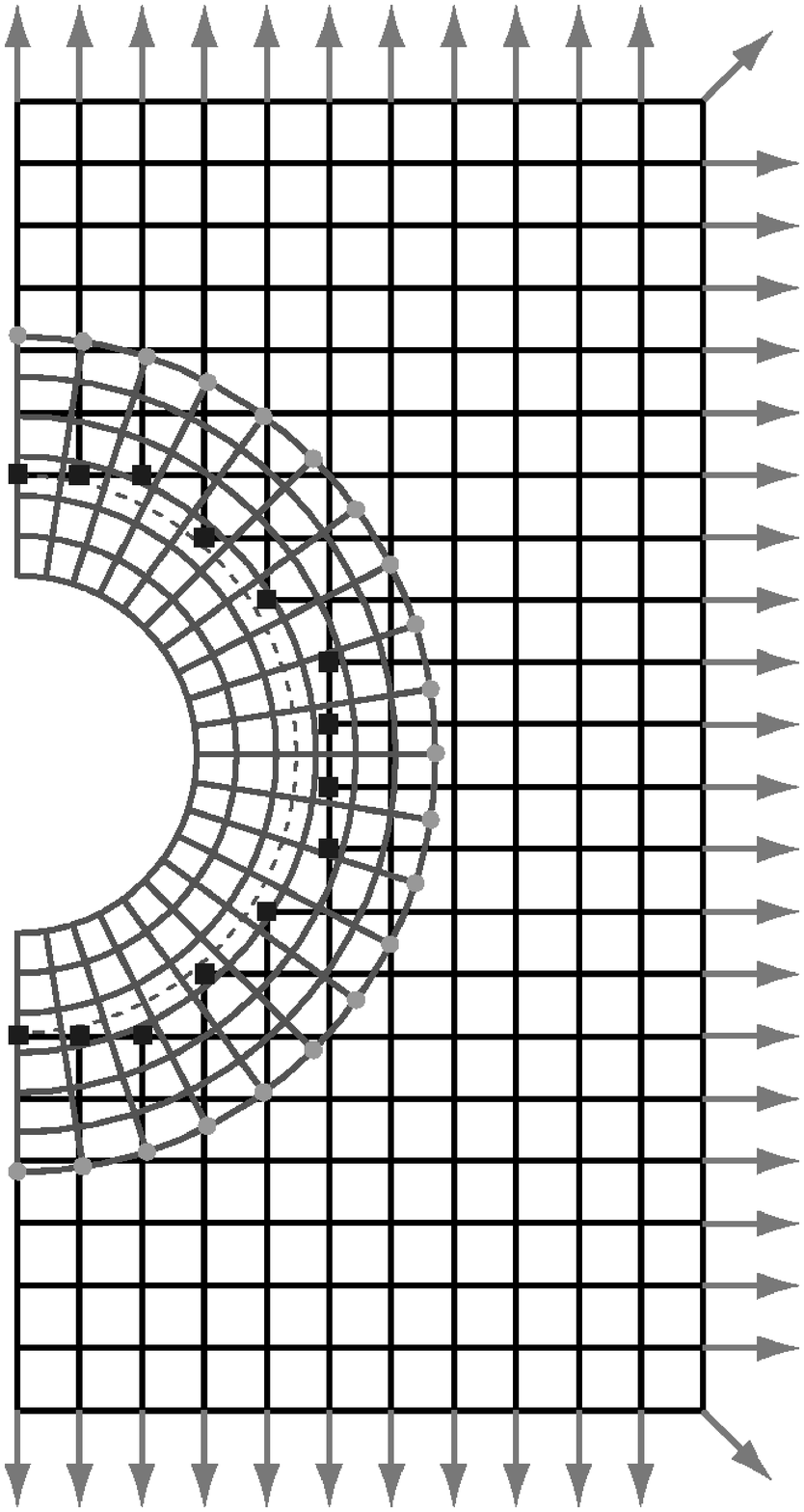}
\includegraphics*[height=6cm]{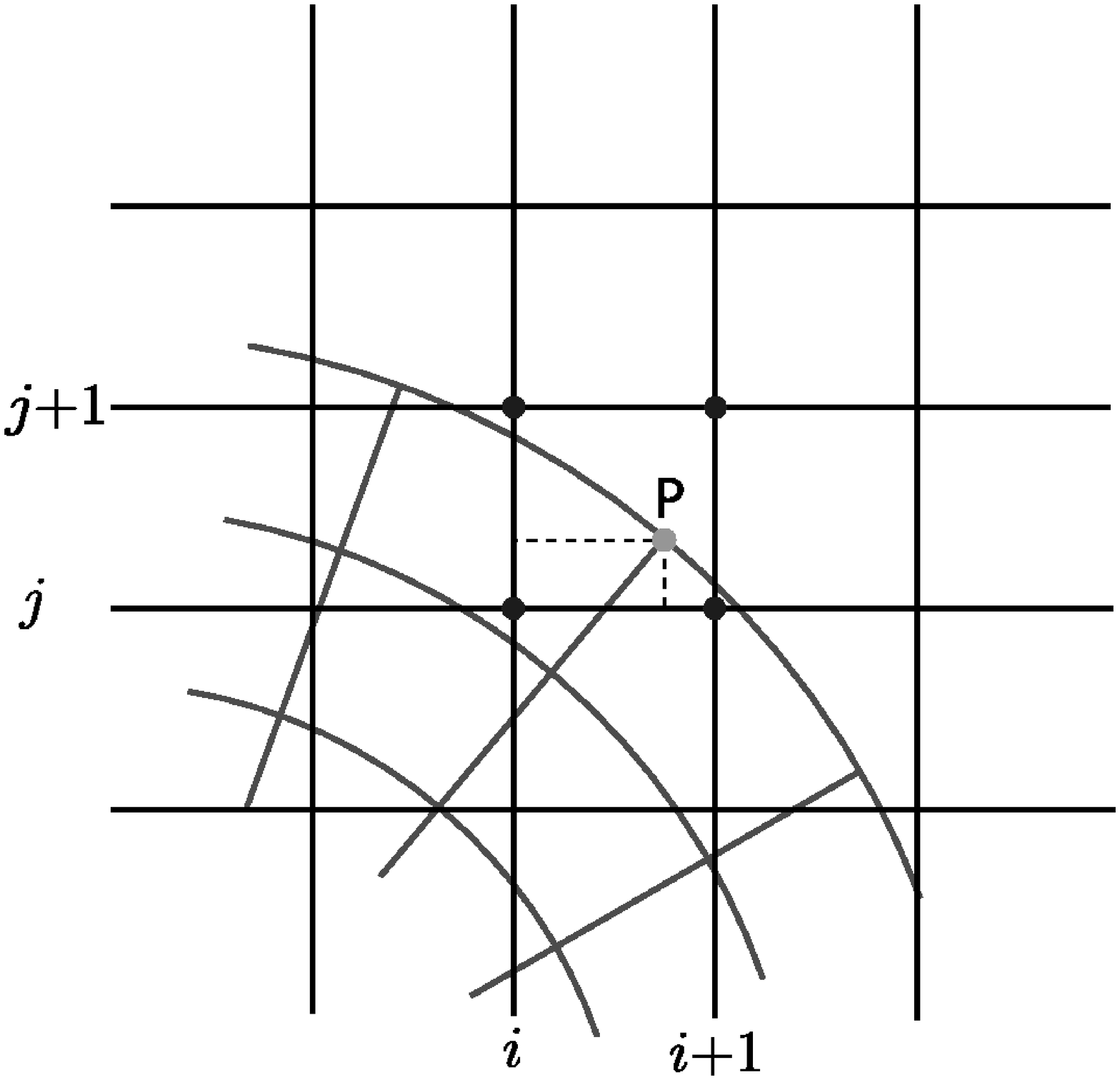}
\caption{The first figure shows the overlapping grids for the
axisymmetric wave equation on a boosted black hole background.  The
spherical grid is used to excise the black hole from the computational
domain.  The dashed line represents a $r'={\rm conts.}$ curve on the
spherical patch.  This is where the cylindrical coordinate system
terminates.  The arrows on the outer boundary indicate how the unit
normal is chosen at each boundary grid point where boundary data must
be specified.  The circles and the square are used to mark points of
the spherical and cylindrical grid respectively which are updated
through interpolation.  As shown in the second figure, the value
of the fields on the grid point $P$ of the spherical grid are computed
by interpolating the values of the fields on the four neighboring
points.  }
\label{Fig:Overlapping}
\end{center}
\end{figure}

As mentioned in Section~\ref{Sec:Overview_excision}, the method
of overlapping grids gives a natural method for solving finite
difference problems on multiple domains.  For the boosted black 
hole,
we use the cylindrical grid as our base grid and introduce the
spherical grid adapted to the inner boundary (event horizon).  The two
grids overlap as shown in Fig.~\ref{Fig:Overlapping}.

The spherical grid requires boundary data at $r =
r_{\rm max}$, which does not constitute a physical boundary in this
problem.  Here the data is computed by interpolating the values of the
field from the cylindrical grid.  Similarly, the values of the fields
at the grid points of the cylindrical grid near the excision region,
which lack a neighboring point in a coordinate direction (these points
are marked with a square in Fig.~\ref{Fig:Overlapping}), are also updated
via interpolation.  In this work we used second order Lagrangian
interpolation, which, for a scalar quantity, is given by
\begin{eqnarray*}
&&f_{\rm Int}(x_i+a \Delta x, y_j + b \Delta y) = (1-a)(1-b)f_{ij} \\
&&+(1-a)b f_{i,j+1} + a(1-b) f_{i+1,j} + ab f_{i+1,j+1}
\end{eqnarray*}
where $0\le a, b < 1$.  Higher-order interpolation stencils may also
be used, though for the cases examined here, improvements in the
resulting solutions are slight, resulting in no increase in the order
of accuracy.

The boosted cylindrical and co-moving spherical coordinate systems 
are related by
\begin{eqnarray}
\bar{t} &=& t'\\
\bar{\rho} &=& r' \sin\theta'\nonumber\\
\bar{z} &=& \gamma^{-1} r' \cos\theta' - \beta t'\nonumber\\
\bar{\phi} &=& \phi',\nonumber
\end{eqnarray}
and the inverse transformation
\begin{eqnarray}
t' &=& \bar{t} \\
r' &=& \sqrt{\bar{\rho}^2 + \gamma^2 (\bar{z} + \beta \bar{t})^2}\nonumber\\
\theta' &=& \tan^{-1}\left(\frac{\bar{\rho}}{\gamma(\bar{z}+\beta
\bar{t})} \right)\nonumber\\
\phi' &=& \bar{\phi}.\nonumber
\end{eqnarray}
The evolved fields are not scalar quantities, but components of a 1-form.
So, in addition to the coordinate
transformation between the two coordinate systems,
the communication of the values of the fields requires the use of the
transformation law of 1-forms.  In this case we have
\begin{eqnarray}
\bar{T} &=& T' + \gamma \beta \cos\theta' R' -
\gamma\beta\frac{\sin\theta'}{r'} \Theta' \\
\bar{P} &=& \sin\theta'
R' + \frac{\cos\theta'}{r'} \Theta'\nonumber\\
\bar{Z} &=& \gamma\cos\theta'R'
-\gamma\frac{\sin\theta'}{r'}\Theta',\nonumber
\end{eqnarray}
and 
\begin{eqnarray}
T' &=& \bar{T} - \beta \bar{Z}\\
R' &=& \sin\theta' \bar{P} +
\frac{\cos\theta'}{\gamma} \bar{Z}\nonumber\\
\Theta' &=& r'\cos\theta' \bar{P}
- r' \frac{\sin\theta'}{\gamma} \bar{Z},\nonumber
\end{eqnarray}
where $(\bar{T},\bar{P},\bar{Z})$ and
$(T',R',\Theta')$ are the fields on the cylindrical and spherical grids,
respectively.

\subsubsection{Discretization on the axis}

The discretization of the system in boosted cylindrical coordinates in
the interior and at the outer boundary is done according to 
Eqs.~(\ref{Eq:DWEgen2})--(\ref{Eq:DWEgen3}) 
where the components $\gamma^{\bar\mu\bar\nu}$ are
given in (\ref{Eq:gamma_cyl_boosted}).  On the axis of symmetry we use 
\begin{eqnarray*}
\partial_{\bar{t}} \bar{T} &=& \left( \tilde{\gamma}^{\bar{t}\bar{z}}
D^{(\bar{z})} \bar{T} + 2 \tilde{\gamma}^{\bar{t} \bar{\rho}}
\bar{T} + D^{(\bar{z})} (\tilde{\gamma}^{\bar{t} \bar{z}}
\bar{T}) + 2 \tilde{\gamma}^{\bar{\rho} \bar{\rho}}
D_+^{(\bar{\rho})} \bar{P} + \right.\\
&&\left.2 \tilde{\gamma}^{\bar{\rho} \bar{z}}
\bar{Z} + D^{(\bar{z})} (\tilde{\gamma}^{\bar{z} \bar{z}}
\bar{Z}) + \partial_{\bar{t}} \tilde{\gamma}^{\bar{t} \bar{t}} \bar{T}
+ \partial_{\bar{t}} \tilde{\gamma}^{\bar{t} \bar{z}} \bar{Z}\right)
/(-\tilde{\gamma}^{\bar{t} \bar{t}}).
\end{eqnarray*}

Similarly, the discretization of the system in co-moving spherical
coordinates in the interior and on the event horizon is done according to
Eqs.~(\ref{Eq:DWEgen2})--(\ref{Eq:DWEgen3}), where the components of
$\gamma^{\mu'\nu'}$ are given in (\ref{Eq:gamma_spher_comoving}).  On the axis
of symmetry ($\theta = 0$ and $\theta = \pi$) we use 
\begin{eqnarray*}
\partial_{t'} T' &=& \left( \tilde{\gamma}^{t'r'}
D^{(r')} T' + D^{(r')}(\tilde{\gamma}^{t'r'}T') \pm
2 \tilde{\gamma}^{t' \theta'}
T' + \right.\\
&&\left.D^{(r')} (\tilde{\gamma}^{r'r'}
R') + 2 \tilde{\gamma}^{\theta'\theta'}
D_{\pm}^{(\theta')} \Theta' \right)
/(-\tilde{\gamma}^{t' t'}).
\end{eqnarray*}

\subsubsection{Boundary conditions}

Boundary conditions in maximally dissipative form are given at the
outer boundary of the cylindrical grid in the directions indicated in
Fig.~\ref{Fig:Overlapping}.  In the boosted case, instead of
overwriting the right hand side at the boundary according to Olsson's
prescription, we overwrite the solution itself.  The reason for doing
so, is that it avoids the tedious task of computing time
derivatives of the boundary data.

It is interesting to notice that the outer boundary of the cylindrical
grid could become at some points purely inflow ($s_{\pm} > 0$) for
very large values of $\beta$.  (We exclude the case in which the black
hole is outside the outer boundary.)  At and near these inflow
boundary points the system is only strongly hyperbolic and the energy
method fails to give the correct boundary conditions.  As it is
pointed out in~\cite{CalSar}, applying maximally dissipative boundary conditions
to strongly hyperbolic systems can lead to an ill posed IBVP.
Where the boundary is purely inflow we give data to the two incoming
fields.  Our numerical experiments (Sec.~\ref{Sec:NumExp}) indicate
that the scheme is stable.

\subsubsection{Artificial dissipation}

Whereas the single grid schemes do not require any artificial
dissipation, it is known that overlapping grids require explicit
dissipation for stability~\cite{OlsPet}.  To the right hand side of
the discretized system a term of the form~\cite{KreOli}
\begin{equation}
Q_d u = -\sigma \left( h_1^3 (D^{(1)}_+D^{(1)}_-)^2 +  h_2^3
(D^{(2)}_+D^{(2)}_-)^2\right) u
\end{equation}
is added.  This dissipative operator is modified near the outer and
inner boundary, as was done in~\cite{CalLehNeiPulReuSarTig}.  Near and
on the axis of symmetry dissipation is computed exploiting the
regularity conditions of the fields.  As this dissipation has a
five-point stencil, we find that the long-term behavior of the code
is improved in some cases by interpolating two points at all
inter-grid boundaries.

\subsubsection{Choice of Courant factor}

\begin{figure}[ht]
\begin{center}
\includegraphics*[height=6cm]{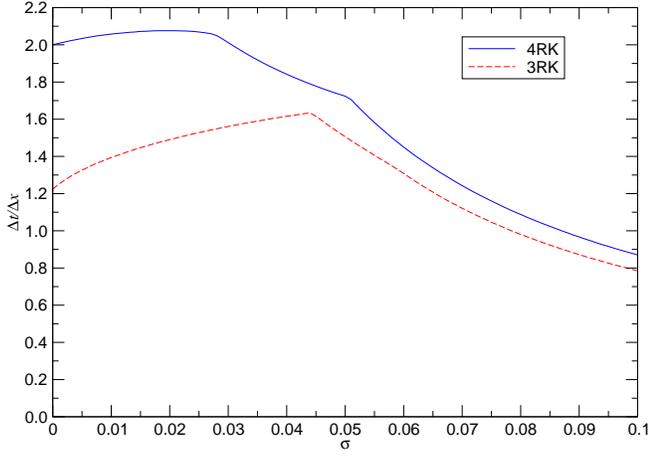}
\caption{The Courant limits for the 2D first order wave equation
$\partial_t u_0 = \partial^i u_i$, $\partial_t u_i = \partial_i u_0$
for fourth order (4RK) and third order Runge-Kutta (3RK).  The
calculation assumes no boundaries and second order centered difference
operators for the approximation of the spatial derivatives.  The value
of $\sigma$ represents the amount of artificial dissipation, $-\sigma \sum_{i=1}^2
h_i^3 (D^{(i)}_+D^{(i)}_-)^2 u$, added to the rhs of the equations.}
\label{Fig:Courant}
\end{center}
\end{figure}

The fully discrete system is obtained by integrating the semi-discrete
system with third or fourth order Runge-Kutta.  Whenever explicit
finite difference schemes are used to approximate hyperbolic problems,
the ratio between the time step size $k$ and the mesh size $h = \min
\{ h_i\}$, the {\em Courant factor}, cannot be greater than a certain
value~\cite{CFL}.  This Courant limit is inversely proportional to the
characteristic speeds of the system.  We estimate allowable values
for the Courant factor by examining the 2D wave equation in
first order form, $\partial_t u_0 = \partial^i u_i$, $\partial_t u_i =
\partial_i u_0$.  Assuming second order, centered differencing
for the spatial derivatives, we plot the Courant limits for 
third and fourth order Runge-Kutta as a function of the artificial 
dissipation parameter in Fig.~\ref{Fig:Courant}.

The characteristic speeds in the cylindrical grid are bounded by 1 in
magnitude.  Looking at Fig.~\ref{Fig:Courant}, this would suggest
that one could use a Courant factor as large as 2.0 (using fourth
order Runge-Kutta and ignoring the fact that this is a variable
coefficient problem with lower order terms and with boundaries).
However, in the spherical grid the characteristic speeds along the
axis of symmetry have a magnitude of
\begin{equation}
\sqrt{\frac{1+|\beta|}{1-|\beta|}},
\end{equation}
which is greater than 1 for $\beta \neq 0$.  For example, for $\beta =
0.9$ the characteristic speeds in the spherical grid can be as large
as $\sqrt{19} \approx 4.359$.  In this case a Courant factor larger
than $\approx 0.46$ is likely to lead to numerical instability.

\section{Numerical Experiments}
\label{Sec:NumExp}
 
To our knowledge there are no stability proofs for two dimensional
hyperbolic problems with overlapping grids.  To check the convergence
of our code we must rely on numerical experimentation. 

Let $u(t,\vec x)$ be the exact solution of the continuum problem and
$v^n_{ij}$ the solution of the fully discrete approximation.  If, for any
$t$, as $nk\to t$,
\[
\epsilon^n_{h} \equiv ( h_1h_2\sum_{ij} \| v^n_{ij}- u(nk,\vec
x_{ij})\|^2)^{1/2} = {\cal O}(h^p) + {\cal O}(k^q)
\]
as $k, h \to 0$, the difference scheme is said to be convergent of
order $(p,q)$.  This implies that the overall order of convergence of
the scheme, assuming $k/h=\mbox{const.}$, is 
\begin{equation}
Q \equiv \lim_{h\to 0} \log_2 \frac{\epsilon^n_h}{\epsilon^n_{h/2}} =
\min\{p,q\}
\label{Eq:convergence}
\end{equation}
as $nk \to t$, where $t$ is some fixed time.  To use this equation we
must know an exact solution of the continuum problem.

\begin{figure}[ht]
\begin{center}
\includegraphics*[height=6cm]{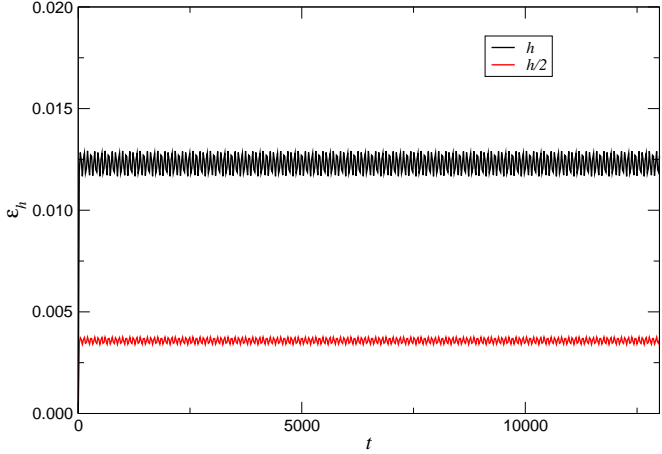}
\caption{
These long term runs done at a fairly course resolution for $\beta = M
= 0$ suggest that the interpolation between the overlapping grids does
not introduce any power law or exponential growth.  Here the exact
solution is given by the the real part of the $\ell=0$ mode with $\omega=2$,
with the in-going Hankel function for the radial variable in
Eq.~(\ref{Eq:Jacwave}).  The dissipation parameter is set to
$\sigma=0.02$.  The ranges of the dependent variables are $0\le \rho
\le 10$, $-10\le z \le 10$, $1 \le r \le 5$, and $0 \le \theta \le \pi$.   
The coarsest resolutions for the cylindrical and spherical grids are
$90\times 170$ and $50\times 68$, respectively.}
\label{Fig:Longconvergence}
\end{center}
\end{figure}

Exact solutions for the scalar wave equation in Minkowski space are well 
known.  To test our overlapping grid system we use spherical 
waves~\cite{Jac} given by
\begin{equation}
\Phi = \sum_{\ell} f_{\ell}(r) P_{\ell}(\cos\theta) e^{-i\omega t},
\label{Eq:Jacwave}
\end{equation}
where $P_{\ell}(\cos\theta)$ are the Legendre polynomials and we
choose $f_{\ell}(r)$ to be the Hankel functions, which
asymptotically represent in- and out-going waves.  We tested the
long-term behavior of our code by evolving an ingoing spherical wave
exact solution, and computing the norm of the error.  These results
are shown in Fig.~\ref{Fig:Longconvergence}.

When an exact solution is not available, which is often the case, the
following standard technique of numerical analysis can be useful.  Let $w$
be an arbitrary function and let us rewrite the partial differential
equation as $L(u) = 0$.  If $w$ is inserted into the equation, in
general, it will produce a non vanishing right hand side,
\begin{equation}
L(w) = F\,.
\end{equation}
Clearly, the modified equation $\tilde{L}(u) \equiv L(u) - F = 0$ has
$w$ as an exact solution and the convergence of the code can be tested
using Eq.~(\ref{Eq:convergence}).

\begin{figure}[ht]
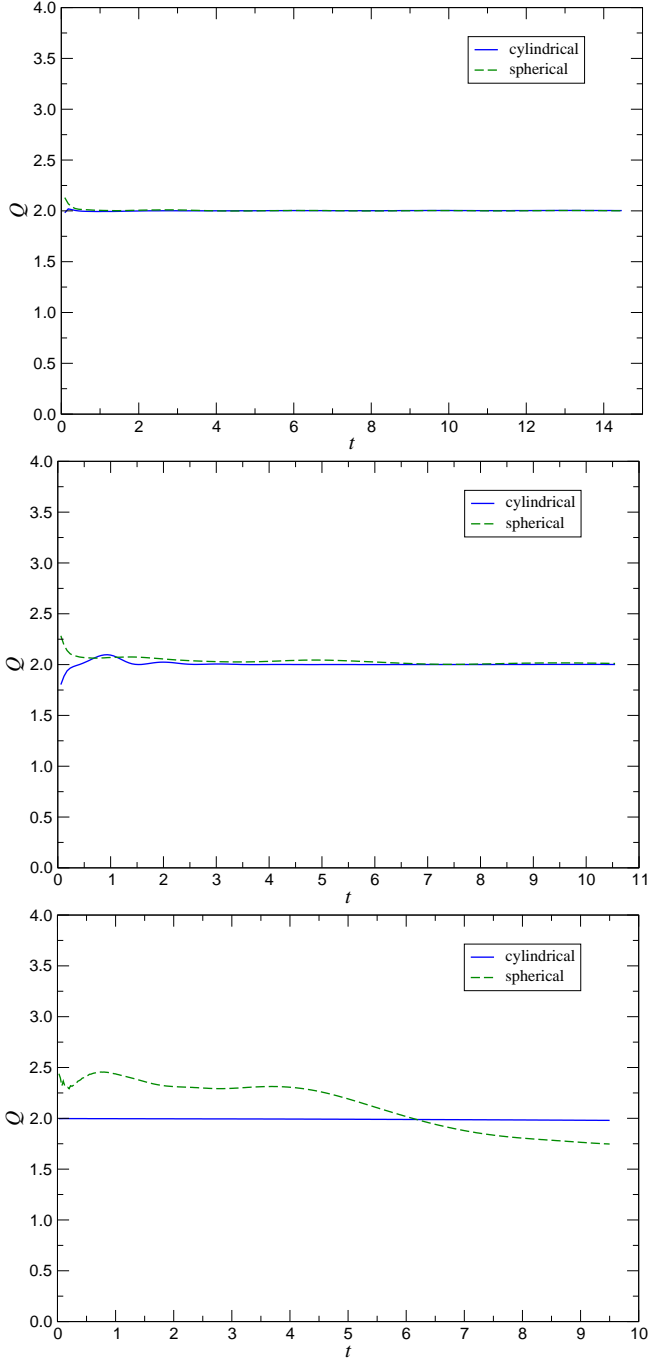

\begin{center}
\includegraphics*[height=6cm]{Q050.eps}
\includegraphics*[height=6cm]{Q075.eps}
\includegraphics*[height=6cm]{Q095.eps}
\caption{In these figures we plot the convergence factor
$Q(t)=\log_2(\epsilon^n_h/\epsilon^n_{h/2})$ as a function of $t=nk$
for a sufficiently small value of the spacing $h$.  The convergence
test was carried out with the modified system (\ref{Eq:WEmodif2}) for
boosting parameters $\beta = -0.5$, $-0.75$, and $-0.95$, from top to
bottom, suggesting that the equations are correctly implemented and
that the overall scheme is second order accurate.  The resolutions
used in the cylindrical and spherical grid are $256\times 512$,
$128\times 384$ and $512\times 1024$, $256\times 768$.  The domain
extends from $-10M$ to $+10M$ in the $\bar z$ direction and up to
$+10M$ in the $\bar \rho$ direction.  The spherical patch covers the
region $2M\le r' \le 3M$.  We used a Courant factor of $1.15$, $0.75$,
and $0.32$, and a dissipation parameter of $\sigma = 0.02$.  The
evolution is stopped just before the spherical grid touches the outer
boundary of the cylindrical grid.  In the $\beta = -0.95$ the system
in only strongly hyperbolic at the bottom right corner of the
cylindrical grid.  We found that, to achieve convergence, we must
give data to all fields at this point.}
\label{Fig:Convergence}
\end{center}
\end{figure}

We chose $w(t,r,\theta) = \sin(t+r) \cos(n\theta)$, where
$\{t,r,\theta\}$ are the spherical coordinates of the rest frame and
$n$ is an integer. This is an exact solution of
\begin{equation}
\nabla_{\mu}\nabla^{\mu} w - F = 0\,.
\end{equation}
where $F$ is given by
\begin{eqnarray}
F &=& \frac{\cos n \theta}{r^2} \left( 2r\cos(t+r) - n^2 \sin(t+r) \right) \\
&&-n\frac{\sin n\theta}{r^2 \sin\theta} \cos\theta \sin(t+r)\,.\nonumber
\end{eqnarray}
Both $w$ and $F$ are scalar quantities.
The evolution equation (\ref{Eq:WEgen2}) is modified according to
\begin{eqnarray}
\partial_t T &=& -\left(\gamma^{ti}\partial_i T +
\partial_i(\gamma^{it}T)+ \partial_i (\gamma^{ij}d_j) \right.
\label{Eq:WEmodif2}\\
&&\left. +
\partial_t \gamma^{tt}T + \partial_t
\gamma^{tj}d_j - \sqrt{-g}F \right)/\gamma^{tt},
\nonumber
\end{eqnarray}
where $g = \det (g_{\mu\nu})$.  On the axis of symmetry we use the
limits $\lim_{\theta\to 0}\frac{\sin n\theta}{\sin\theta} = n$ and
$\lim_{\theta\to \pi}\frac{\sin n\theta}{\sin\theta} = (-1)^{n+1}n$.
The results of our convergence tests for different values of the boost
parameter and $n=2$ are summarized in Fig.~\ref{Fig:Convergence}.  
Movies are also available~\cite{movies}.

\section{Conclusion}

Systems with moving boundaries arise in a variety of situations, and
their solution typically involves introducing coordinates adapted to
the boundaries.  These may be either a single, global coordinate
system, such as those used in binary black hole
evolutions~\cite{AlcBruPolSeiTak,AlcBenBruLanNerSeiTak} that keep the
black holes and the outer boundary at fixed coordinate positions, or,
as advocated here, multiple coordinate patches.  Whichever approach is
adopted, fixing coordinates to the boundaries allows one to
unambiguously specify proper boundary conditions, as required
for well-posed problems.  Moreover, boundaries at fixed grid
coordinates eliminate the need for extrapolated data at points that
emerge from a moving boundary.

In our model problem we have evolved an axisymmetric scalar field on a
boosted Schwarzschild background.  We used a cylindrical coordinate
patch with respect to which the outer boundary is fixed, and we
introduced an overlapping spherical coordinate patch co-moving with
the hole.  At any given time in the co-moving coordinate system the
location of the inner boundary (the horizon) corresponds to a $r={\rm
const.}$ surface.  This surface can be represented exactly on the
numerical grid and allows one to smoothly excise a large volume of
spacetime, much larger than that permitted by cubical excision.  Our
numerical implementation made use of overlapping grids, where
different but equivalent problems are solved on separate grids.  To
communicate data between grids we used interpolation on all fields.

The discrete version of the energy method, based on differencing
operators that satisfy the summation by parts property, has
demonstrated to be particularly effective for the construction of a
stable discretization scheme on the axis of symmetry, and for the
identification of discrete boundary conditions.  The stability proofs,
which hold on individual grids, cannot be immediately extended to the
overlapping grid scheme, due to the interpolation of data from one
grid to another.  We note, however, that it may be possible to define
orthogonal projection operators for the interpolation that could allow
to analytically demonstrate numerical stability~\cite{POlsson}.  This
is a question of active interest.  Nevertheless, our numerical tests
indicate that our scheme is convergent, even for very high values of
the boost parameter, and does not suffer from long term power-law or
exponential growth in the error.

The model problem that we presented in this work is primarily intended
as a proof of concept, and several avenues of research remain to be
explored.  Foremost might be the addition of the Einstein equations to
the system for a dynamic black-hole spacetime, where the locations of
the black hole singularity and event horizon are not a priori known.
Depending on the dimensionality of the problem, one or more coordinate
patches adapted to the inner boundary would have to be generated
during evolution, along with the relationship between the various
coordinate systems.  By monitoring the characteristic speeds on the
excision boundary (with respect to the coordinate system adapted to
that boundary), one can guarantee its purely outflow properties, an essential
requirement of excision.

Although alternative numerical approaches may be possible, the
overlapping grid method has struck us for its strength and its
simplicity.  Owing to its flexibility in representing smooth, time
dependent boundaries, we believe that this technique, or a similar one,
will play a significant role in the solution of the binary black hole
problem.

\section*{Acknowledgments}
We thank Olivier Sarbach for many valuable discussions during this
work, and for sharing early results on summation by parts for the two
dimensional wave equation in axisymmetry.  We thank Luis Lehner for
sharing results of an earlier investigation of cubical excision in the
Kerr spacetime, as well as for encouragement in the beginning phases
of this project.  We acknowledge Manuel Tiglio for interesting
discussions and for providing Ref.~\cite{BasReb}.  We also thank
N.~Andersson, M.~Choptuik, S.~Liebling, P.~Olsson, F.~Pretorius,
J.~Pullin, O.~Reula, and S.~Teukolsky for helpful discussions.  We
thank L.~Lehner and J.~Pullin for comments on an earlier draft of this
work and S.~Ou for help with visualization.  We acknowledge the
hospitality of the Kavli Institute of Theoretical Physics at UCSB
(DN), the Max-Planck-Institut f\"ur Gravitationsphysik
(Albert-Einstein-Institut) and the University of Southampton (GC),
where part of this research was done.  This research was supported in
part by the National Science Foundation under grants PHY-9907949 to
the KITP and PHY-0244335 and INT-0204937 to LSU, and the Horace Hearne
Jr. Institute for Theoretical Physics.

\section*{Appendix}

To gain some insight into the limitations of cubical excision and the
consequent need for a smooth excision boundary, we consider the
analytic Schwarzschild and Kerr solutions~\cite{Sch,Leh}.  
We employ the commonly
used Cartesian Kerr--Schild coordinates, which are smooth across the
horizon, and write the metric as
\[
g_{\mu\nu} = \eta_{\mu\nu} + 2 H \ell_\mu\ell_\nu\,,
\]
where $\eta_{\mu\nu}$ is the Minkowski metric, $H$ is a scalar,
\[
H = \frac{Mr}{r^2 + a^2\cos^2\theta}\,,
\]
and $\ell_\mu$ is a null vector,
\[
\ell_\mu = \left(1, \frac{rx + ay}{r^2 + a^2},
                 \frac{ry - ax}{r^2 + a^2}, \frac{z}{r}
           \right)\,.
\]
The parameter $M$ represents the mass of the black hole and $J = aM$
is the total angular momentum.  The spheroidal coordinates $r$ and
$\theta$ are given by
\[
\cos\theta = \frac{z}{r}
\]
\[
r^2 = \frac{1}{2}(\rho^2 - a^2) + \sqrt{\frac{1}{4}(\rho^2 - a^2)^2 + a^2z^2},
\]
where $\rho^2 = x^2 + y^2 + z^2$.

\begin{figure}[ht]
\begin{center}
\includegraphics*[height=6cm]{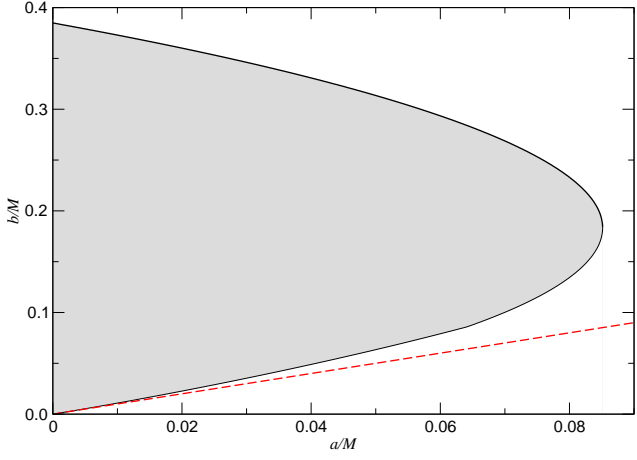}
\caption{This figure indicates the limitations of cubical excision in
the Kerr spacetime in rectangular Kerr--Schild coordinates.  
We assume that the
excision cube is centered on the hole, and that the faces of the cube
are at $\pm b$.  (See description in text.)
Values of $b$ for which an inner boundary has no
incoming modes, and thus a candidate for an excision boundary, are
indicated by the shaded region in the figure.  The structure of the
Kerr spacetime results in both maximum and minimum limits to the size
of the excision cube.  We find that, in this particular coordinate
system, cubical excision for Kerr is well-defined only for very small
spin parameters, $a\lesssim 0.0851M$, where $a=J/M$.  For values of
$b$ below the dashed line, the excision cube intersects the ring
singularity.}
\label{fig:Kerr_cubical_excision}
\end{center}
\end{figure}

We center a cube of side length $L = 2b$ on the black hole, $x^i \in
[-b,b]$.  In order to excise this region from the computational
domain, we must ensure that its boundary is purely outflow, i.e., that
no information can enter the computational domain.  To determine the
allowed values of $b$, we calculate the characteristic speeds on each
face of the cube and check that the inequality $s^n_\pm \le 0$, where
$\vec n$ is the outward unit normal to the boundary, is satisfied.
The Schwarzschild solution is obtained by setting $a=0$, and the
calculation gives~\cite{Sch} $0<b \le 2\sqrt{3}/9 M \approx 0.385M$.
The calculations for Kerr ($a\neq 0$) are more involved, and we
present our numerically generated results in
Fig.~\ref{fig:Kerr_cubical_excision}.  We find that because of the
ring singularity ($\rho = a$, $z=0$), in addition to a maximum size
for the excision cube, there is also a {\em minimum} size.  In
addition, we notice that no cubical excision is possible for $a
\gtrsim 0.0851M$.  This is a severe constraint on the spin parameter,
and precludes cubical excision for interesting values of spin.  We
note, however, that this limitation is coordinate dependent and that
it might be possible to choose coordinates in which cubical excision
may be done for higher values of $a$.


\end{document}